\documentclass[sigconf,nonacm,authorversion]{acmart}
\usepackage{listings}
\usepackage{booktabs}
\usepackage{csquotes}
\usepackage{tabularx}
\usepackage{graphics}
\usepackage{tikz}
\usetikzlibrary{shapes.geometric,arrows.meta,positioning,calc}
\usepackage[vskip=1em,font=itshape,leftmargin=2em,rightmargin=2em]{quoting}

\lstdefinestyle{format}{
    basicstyle=\ttfamily\footnotesize,
    breakatwhitespace=false,
    breaklines=true,
    captionpos=b,
    numbersep=8pt,
    tabsize=2,
    columns=fullflexible
}

\AtBeginDocument{%
  }

\setcopyright{acmlicensed}
\copyrightyear{2025}
\acmYear{2025}
\acmDOI{XXXXXXX.XXXXXXX}

\acmConference[CHI '27]{ACM Conference on Human Factors in Computing Systems}{May 2027}{Pittsburgh, PA, USA}
\acmISBN{978-1-4503-XXXX-X/25/10}

\begin{document}

\title{Atlas of Human-AI Interaction: An Interactive Knowledge Graph for Research Sensemaking at Scale}

\renewcommand{\shorttitle}{Atlas of Human-AI Interaction}

\begin{abstract}
The growth and fragmentation of scientific literature make it increasingly difficult for researchers to synthesize findings and build on prior work. This paper presents a framework for organizing empirical findings to support comparison and exploration across papers. The framework combines LLM-powered claim extraction, entity resolution, taxonomy generation, and knowledge graph construction, retaining links to study contexts and sources. An interactive interface supports navigation among constructs and reported relationships, alongside evidence inspection and collection. We instantiate the framework in the Atlas of Human-AI Interaction, organizing 308,802 claims from 31,741 papers. A descriptive corpus analysis and exploratory user study examine how this representation can support research sensemaking and identify opportunities for refinement. The framework and interface offer an initial approach to computational literature synthesis, with potential for adaptation to other empirical domains.
\end{abstract}

\ccsdesc[500]{Human-centered computing~Human computer interaction (HCI)}
\ccsdesc[500]{Computing methodologies~Artificial intelligence}
\ccsdesc[500]{Information systems~Information retrieval}

\keywords{Human-AI Interaction, Causal Knowledge Graph, Research Synthesis, Literature Navigation, Meta-Science}

\author{Chayapatr (Pub) Archiwaranguprok}
\affiliation{%
  \institution{MIT Media Lab, Massachusetts Institute of Technology}
  \city{Cambridge}
  \state{Massachusetts}
  \country{USA}
}
\email{pub@mit.edu}

\author{Rachel Poonsiriwong}
\affiliation{%
  \institution{MIT Media Lab, Massachusetts Institute of Technology}
  \city{Cambridge}
  \state{Massachusetts}
  \country{USA}
}
\email{rachelpo@media.mit.edu}

\author{Awu Chen}
\affiliation{%
  \institution{MIT Media Lab, Massachusetts Institute of Technology}
  \city{Cambridge}
  \state{Massachusetts}
  \country{USA}
}
\email{awuchen@media.mit.edu}

\author{Pattie Maes}
\affiliation{%
  \institution{MIT Media Lab, Massachusetts Institute of Technology}
  \city{Cambridge}
  \state{Massachusetts}
  \country{USA}
}
\email{pattie@media.mit.edu}

\author{Pat Pataranutaporn}
\affiliation{%
  \institution{MIT Media Lab, Massachusetts Institute of Technology}
  \city{Cambridge}
  \state{Massachusetts}
  \country{USA}
}
\email{patpat@media.mit.edu}

\begin{teaserfigure}
  \includegraphics[width=\textwidth]{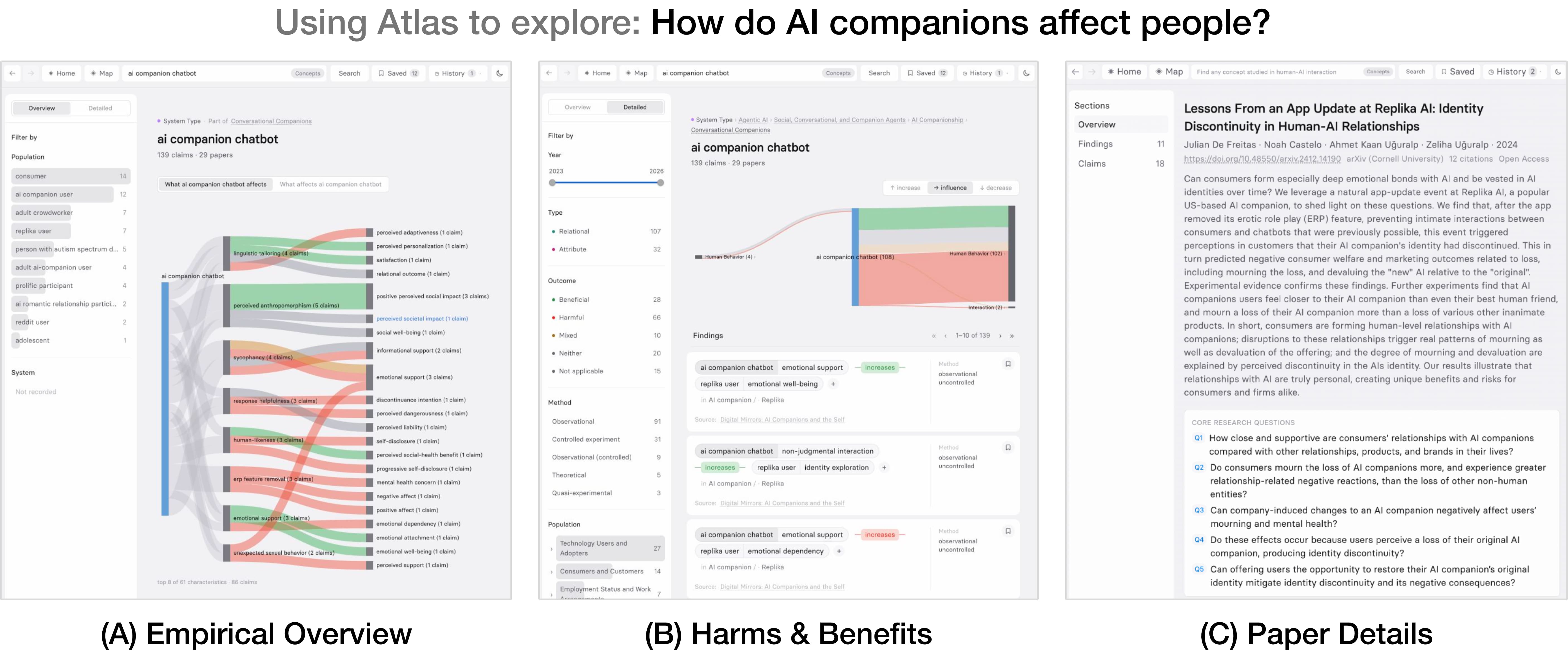}
\caption{\textbf{Tracing evidence about companion AI with Atlas.} (A) The AI companion chatbot overview connects system characteristics with outcomes such as emotional well-being, dependency, and responses to system changes. (B) The detailed view exposes taxonomy structure and filters for outcomes, methods, populations, and systems. (C) Claim cards retain an expanded source passage alongside related findings about Replika users. The paper view shows \emph{Lessons From an App Update at Replika AI: Identity Discontinuity in Human-AI Relationships} and its study designs, including surveys, an archival Reddit analysis, and controlled experiments. The sequence connects concept-level relationships to empirical findings and their provenance.}
\Description{Three portrait screenshots illustrate a path through Atlas from an AI companion chatbot overview to individual claims and a source paper. Panel A shows the overview filtered to Replika users, with ERP feature removal selected. Panel B shows Outcome: Harmful and Method: Quasi-experimental filters, with an expanded finding about negative Reddit posts after an app update. Panel C shows the Replika app-update paper with its abstract, research questions, and five study designs. All panels use light mode and enlarged interface text.}
    \label{fig:atlas-companion-teaser}
  \label{fig:teaser}
\end{teaserfigure}

\maketitle

\section{Introduction}

\begin{quoting}
The complexity of challenges we face may soon outpace our ability to solve them ... So a core challenge for organizations tackling important challenges: Getting smarter at getting smarter... -- \textbf{Douglas C. Engelbart}
\end{quoting}

\noindent Scientific literature now grows faster than researchers can read it. The volume of published research has increased exponentially over the past three centuries~\cite{bornmann2015growth, bornmann2021growth}, with indexed articles roughly 47 percent more numerous in 2022 than in 2016, outpacing any growth in the number of practicing scientists~\cite{hanson2024strain}, and no empirical sign of a plateau~\cite{bornmann2015growth, parolo2015attention}. In human-AI interaction, over ten thousand papers appeared in 2025, an average of 30.2 per day\footnote{See Section~\ref{sec:pipeline}}, which places the field's annual output beyond what any individual researcher can reasonably make sense of it.

The consequences of this expansion are both epistemic and infrastructural. For individual researchers, the problem is no longer simply finding papers, but establishing what is already known across increasingly technical and fragmented literatures whose related findings may be expressed through different constructs, methods, and disciplinary vocabularies. Researchers unable to survey this landscape risk duplicating rather than building upon prior work, producing avoidable waste that accumulates across stages of the research process and slows cumulative scientific progress ~\cite{chalmers2009avoidable,glasziou2018research,puljak2023definition}. The same growth also places pressure on the institutions responsible for evaluating knowledge. CHI's recent restructuring of its papers review process, for example, introduced a rubric-based Assisted Desk Reject stage in which subcommittee chairs and associate chairs assess submissions before they proceed to full review~\cite{chi2026adr}. This additional screening stage illustrates how the field's growth changes not only what scholars must read, but also how research communities allocate limited expert attention. Comparable pressures appear in evidence synthesis. During COVID-19, review production expanded so rapidly that researchers described an \textit{evidence-synthesis infodemic}, in which the volume of reviews outpaced their quality and made it difficult to determine what was reliably known ~\cite{abbott2022characteristics,haddaway2020eight}. Poorly conducted syntheses can consequently misdirect policy and erode trust in evidence-based practice~\cite{haddaway2020eight}.

Literature synthesis, the central mechanism by which this knowledge is consolidated, is structurally constrained at two levels. At the level of the field, work concentrates within familiar subfields while insights of similar form remain unrecognized elsewhere under different constructs. Analyzing 45 million papers across six decades, ~\cite{park2023papers} find scholarship progressively less likely to push science in new directions and attribute this decline to a narrowing in the use of prior knowledge. At the level of the individual review, the labor has long outstripped capacity, with the daily volume of new evidence having rendered unsystematic narrative synthesis inadequate over a decade ago~\cite{bastian2010seventy}. Together these describe the condition Engelbart foresaw, in which what we know accumulates faster than our means of making sense of it.

Recent work has begun adopting artificial intelligence and machine learning methods to support research sensemaking~\cite{xu2024large, totowards}. Following the framing in~\cite{kang2023synergi}, this work divides into two facets. Top-down synthesis generates structure over a literature through facets, taxonomies, and thematic overviews~\cite{gurcan2021mapping, suominen2016map,Shahaf2012MetroMO,zhang2018taxogen}, including knowledge graph infrastructures that represent scholarly contributions in symbolic forms~\cite{jaradeh2019open, auer2020improving}, though such structures are hard to personalize~\cite{kang2023synergi} and scale poorly when they depend on manual curation~\cite{nicholson2022expanding}. Bottom-up synthesis instead assists researchers through the pipeline itself, from acquisition through summarization. For example, PaperQA2 matched or exceeded expert performance on retrieval, summarization, and contradiction detection~\cite{skarlinski2024language}, while OpenScholar achieved expert-level citation accuracy across 45 million open-access papers~\cite{asai2026synthesizing}, and interactive systems support reading and organization more directly~\cite{kang2022threddy, kang2023synergi, fok2023scim}.

\label{sec:intro-gap}
These approaches offer complementary support for synthesis. Topical and citation-based overviews reveal broad patterns, while tools operating over retrieved or selected papers help readers examine evidence around a developing question. Advances in language models and structured extraction create an opportunity to connect these scales through computational representations of the constructs and relationships reported across a large literature. We investigate this opportunity through a synthesis pipeline designed to capture findings from empirical research. The pipeline extracts findings from full-text papers as typed claims, resolves the constructs they name into a shared vocabulary, and organizes those constructs into taxonomies. It combines these representations into a knowledge graph linking findings across papers while retaining their study contexts and sources.

We instantiate the pipeline in the \textbf{Atlas of Human-AI Interaction}, organizing $308{,}802$ claims from $31{,}741$ papers across six facets. To support exploration of this corpus, we draw on interaction design and cognitive science to define three design goals centered on source traceability, progressive disclosure, and reader control. We implement these goals in an interactive interface that supports navigation among claims, concepts, and corpus-level patterns, alongside evidence collection and conversational synthesis. Together, the pipeline and interface provide a means to investigate how a shared representation of empirical findings can support research sensemaking at scale. The paper contributes the following:

\begin{itemize}
    \item A synthesis pipeline that integrates structured extraction, entity resolution, taxonomy generation, and knowledge graph construction to organize empirical findings across papers.
    \item Atlas, a research sensemaking interface grounded in interaction design and cognitive science, supporting exploration of reported relationships, inspection of source evidence, and collection of claims for synthesis.
    \item An instantiation of the pipeline in human-AI interaction research, producing $308{,}802$ claims from $31{,}741$ papers, together with a descriptive analysis of research themes, reported relationships, and evidentiary bases in the resulting corpus.
    \item A task-based study with ten researchers and students, providing initial evidence of how Atlas supports literature exploration and comparison, alongside difficulties that inform further design.
\end{itemize}

We investigate this approach through system development, descriptive corpus exploration, and a user study. Section~\ref{sec:pipeline} presents the synthesis pipeline and its instantiation in human-AI interaction research. Section~\ref{sec:design} introduces the design goals and interface. Section~\ref{sec:analysis} describes the resulting corpus and illustrates how connected findings support exploration across studies. Finally, a task-based study with ten researchers and students examines literature synthesis and research-question development through descriptive task outcomes and qualitative accounts of participants' experiences.

Organizing empirical findings around shared constructs and explicit relationships makes it possible to query what studies report about a concept and examine the supporting evidence. Through this representation, we aim to help researchers and practitioners navigate human-AI interaction research and identify questions for further investigation. We offer an initial framework for research sensemaking infrastructure that can develop alongside the literature, contributing to the broader effort to support knowledge synthesis in the age of AI.

\section{Background and Related Works}
\subsection{Computational and AI-Assisted Literature Synthesis}

Generative AI has expanded literature synthesis toward retrieving and
combining evidence around research questions
~\cite{skarlinski2024language,asai2026synthesizing}.
Source-backed generation supports inspection of the material underlying
an answer~\cite{singh2025ai2scholarqaorganized}, while automated review workflows extend
this assistance to organizing literature and drafting longer accounts
~\cite{agarwal2025litllmtoolkitscientificliterature,wang2024autosurvey}.
These advances support focused synthesis, but assessing the evidence
included in an answer remains distinct from identifying relevant
material outside its scope.

A complementary direction supports researchers in developing the
questions that guide synthesis. Structured cross-paper context can
improve the coherence and breadth of literature-review outlines
~\cite{Palani_2023}, while generated topic hierarchies provide
starting points for further exploration~\cite{katz-etal-2024-knowledge}.
Mixed-initiative workflows let readers revise these structures as their
interests develop~\cite{kang2023synergi}. Connecting extracted evidence,
facets, and taxonomies further supports movement between study details
and broader interpretations~\cite{fok2025facetstaxonomiessynthesesnavigating}.

The scope and granularity of these structures shape what readers can
discover. Overviews built around research threads or retrieved collections
support exploration within a developing inquiry
~\cite{kang2023synergi,katz-etal-2024-knowledge}, while analysis across facets
remains an opportunity for further development~\cite{fok2025facetstaxonomiessynthesesnavigating}.
Broader bibliometric maps expose citation and co-occurrence patterns
~\cite{van2010software}, but these links do not directly encode
the empirical relationships reported within studies.

\subsection{Structured Representations of Scholarly Knowledge}

A growing body of work explores structured representations as a basis
for organizing and synthesizing research, including scholarly knowledge
graphs~\cite{jaradeh2019open}, contribution comparisons
~\cite{Oelen_2020}, and faceted evidence tables~\cite{fok2025facetstaxonomiessynthesesnavigating}.
These approaches raise a shared challenge of balancing representational
detail with the effort required to construct and maintain it.
Fine-grained claim graphs capture attributes that qualify entities and
relationships~\cite{magnusson2021extractingfinegrainedknowledgegraphs}, but extending
structured knowledge across a growing literature can require substantial
manual annotation, curation, and relation-specific development
~\cite{nicholson2022expanding}. Maintaining meaningful hierarchies as
research evolves introduces further organizational work
~\cite{kargupta2025taxoadapt}.

Language models offer opportunities to make detailed representations
more practical at scale. Entity matching can draw on language-model
knowledge with less dependence on task-specific annotations
~\cite{peeters2024entitymatchingusinglarge}, while candidate-aware
comparisons improve alignment~\cite{wang2025match}.
Taxonomy generation and expansion similarly benefit from learned
parent and sibling relationships~\cite{shen-etal-2025-unified}
and corpus-grounded refinement of hierarchical structure
~\cite{kargupta2025taxoadapt}. These developments motivate investigating
how extraction, terminology alignment, and hierarchical organization
can work together to support fine-grained synthesis across a broad
corpus.

\subsection{Interactive Interface for Literature Sensemaking}

Literature interfaces shape understanding through the relationships they
make available for inspection. Maps of research trajectories foreground
connections among papers~\cite{Shahaf2012MetroMO}, while thread-based
organization connects passages and references into research narratives
~\cite{kang2022threddy,Palani_2023}. Faceted representations
instead support comparison through shared study characteristics
~\cite{fok2025facetstaxonomiessynthesesnavigating}. Each provides an abstraction that helps readers
organize a literature, while also creating a need to examine the
differences it compresses. Linking these abstractions to source material
supports movement between an emerging interpretation and the evidence
on which it rests~\cite{fok2025facetstaxonomiessynthesesnavigating,fok2024qlarifyrecursivelyexpandableabstracts}.

This interpretation also shapes what researchers seek next, making
literature sensemaking an iterative exchange between organizing evidence
and extending its scope. Revisable research threads
~\cite{kang2023synergi} and exploratory questions
~\cite{10.1145/3654777.3676366} allow readers to adjust their inquiry as
they encounter unfamiliar concepts and connections. Retaining discoveries
across these shifts helps understanding accumulate through exploration
~\cite{kang2022threddy}. Atlas builds on this direction by making
reported relationships between constructs directly navigable, connecting
broader patterns to individual claims and study contexts while allowing
readers to collect evidence for a developing synthesis.

Sensemaking develops as researchers organize evidence and revise their
questions~\cite{kang2023synergi,10.1145/3654777.3676366}. Visual structure
can guide this process: in AI-assisted hypothesis exploration, node-link
diagrams acted as ``guardrails,'' supporting orientation and backtracking
~\cite{10.1145/3698061.3726935}. 

The thread- and question-based systems discussed above organize exploration around a developing inquiry, including when they generate an overview of the selected literature. Fine-grained symbolic representations offer a complementary basis for navigation, making constructs and their reported relationships explicit across papers~\cite{magnusson2021extractingfinegrainedknowledgegraphs,jaradeh2019open}. Atlas investigates how such a representation can support movement from corpus-level structure to individual findings and study contexts.

\section{Synthesis Pipeline and Implementation}
\label{sec:pipeline}

The pipeline organizes empirical findings across papers while preserving the study context of each claim. As discussed in Section~\ref{sec:intro-gap}, connecting findings at this scale requires both a common representation of what studies report and a vocabulary through which related constructs can be retrieved together.

Following corpus acquisition and screening, the pipeline proceeds through four stages. Extraction records each paper's findings as typed claims
(Section~\ref{sec:extraction}). Entity resolution maps the constructs named in those claims to a shared vocabulary (Section~\ref{sec:resolution}). Taxonomy generation organizes this vocabulary into a hierarchy for each facet (Section~\ref{sec:taxonomy}). Finally, graph construction integrates the claims, canonical entities, and taxonomies into a structure that readers can traverse from broad categories to individual findings (Section~\ref{sec:graph}).

Each stage is presented in two parts. \emph{Method} states a procedure
formulated to apply to other empirical literatures. \emph{Instantiation} reports
the parameters, models, and counts it takes for human-AI interaction research.
Another corpus would re-derive them.

\subsection{Corpus Acquisition and Screening}

Corpus construction comprises metadata retrieval, relevance screening, and full-text acquisition. A conjunctive query over titles and abstracts establishes the candidate pool, requiring terms referring to both a participant population and a class of system. A classifier then reads each title and abstract to assess domain relevance and the presence of an empirical study. Each assessment takes one of three values. \emph{Yes} and \emph{no} record a determination, while \emph{unclear} indicates insufficient information in the abstract. Inclusion requires \emph{yes} on both axes. Full text is then retrieved for the retained works through a range of sources.

\textbf{Instantiation.} For the Atlas prototype, candidate works were retrieved from OpenAlex, requiring a human-referent term (human, user, participant, team, people) together with an AI-referent term (AI, artificial intelligence, machine learning, LLM, large language model, chatbot, conversational agent, recommender system, recommendation system). Term lists were expanded from a seed set of papers known to belong in the corpus and refined until the query recovered it. Retrieval was limited to DOI registrant prefixes associated with major scholarly publishers and established preprint servers commonly used in human-AI interaction research. This iteration was limited to open-access works. The query returned $243{,}378$ works in total.

Screening used OpenAI \texttt{gpt-5.6-luna} and retained $40{,}148$ works. The retrieval cascade drew on OpenAlex direct URLs, Unpaywall locations, publisher APIs for arXiv, Springer, Elsevier, IEEE and PubMed Central, and the OSF API. Retrieved documents were parsed to plain text using \texttt{PyMuPDF}, yielding full text for $32{,}470$ works. Retrieval failures included withdrawn resources, erroneous indexed locations, and institutional landing pages without machine-identifiable download links. Figure~\ref{fig:prisma} summarizes corpus construction.

\begin{figure}[t]
\centering
\begin{tikzpicture}[
  x=1mm, y=1mm,
  keep/.style={draw, align=left, inner sep=1.2mm, text width=34mm,
               font=\scriptsize},
  drop/.style={draw, gray!70, align=left, inner sep=1.2mm, text width=25mm,
               font=\scriptsize, text=black!70},
  stage/.style={rotate=90, font=\scriptsize\bfseries, anchor=center,
                text=black!60},
  arr/.style={-{Latex[length=1.5mm]}, thin},
  darr/.style={-{Latex[length=1.5mm]}, thin, gray}
]

\node[keep] (id)  at (0,0)   {\textbf{Records identified}\\OpenAlex query \hfill $n = 243{,}378$};
\node[keep] (scr) at (0,-18) {\textbf{Records screened in}\\relevant and empirical \hfill $n = 40{,}148$};
\node[keep] (ret) at (0,-37) {\textbf{Full text retrieved}\\open access \hfill $n = 32{,}470$};
\node[keep] (inc) at (0,-55) {\textbf{Papers contributing claims}\\$308{,}802$ claims \hfill $n = 31{,}741$};

\node[drop] (d1) at (40,-9)  {excluded at screening\\\mbox{$n = 203{,}230$}};
\node[drop] (d2) at (40,-28) {full text unavailable\\\mbox{$n = 7{,}678$}};
\node[drop] (d3) at (40,-46) {no claim extracted\\\mbox{$n = 729$}};

\draw[arr] (id.south)  -- (scr.north);
\draw[arr] (scr.south) -- (ret.north);
\draw[arr] (ret.south) -- (inc.north);

\draw[darr] (0,-9)  -- (d1.west);
\draw[darr] (0,-28) -- (d2.west);
\draw[darr] (0,-46) -- (d3.west);

\node[stage] at (-22,0)   {Identification};
\node[stage] at (-22,-28) {Screening};
\node[stage] at (-22,-55) {Included};

\end{tikzpicture}
\caption{PRISMA flow of corpus construction. An OpenAlex query identified
$243{,}378$ works, of which $40{,}148$ were screened in, $32{,}470$ retrieved in
full text, and $31{,}741$ yielded the $308{,}802$ claims held in the Atlas.}
\label{fig:prisma}
\end{figure}

\subsection{Structured Extraction}
\label{sec:extraction}

\cite{liu2026humanwrittenpaperagentnativeresearch} describe a “storytelling tax,” in which the conventions of linear narrative leave parts of the research process unreported. This perspective motivates examining the structure embedded in published findings. Identifying the constructs studied, the relationships reported, and their study contexts offers a common basis for comparing findings across papers despite differences in terminology and presentation. Extraction records these findings in a common schema while retaining references to their source passages. It proceeds in two stages. The first enumerates a paper's findings as near-verbatim excerpts or close paraphrases, each with a locator to its source section. The second interprets this enumeration in the context of the full text and represents the findings as typed claims. The two stages separate finding identification from schema assignment. Two considerations inform the schema.

\begin{enumerate}
    \item \textbf{Claim form and relation type}: The schema distinguishes relational, comparative, and attribute claims to capture the different forms in which empirical results are reported. Relational claims distinguish effects, moderation, and mediation~\cite{baron1986moderator}. An \emph{effect} claim records whether $X$ increases, decreases, or has no effect on $Y$. A \emph{moderation} claim identifies a variable that changes this relationship, while a \emph{mediation} claim records a pathway through an intermediate variable. For example, transparency may increase trust, with the relationship varying by prior experience or operating through perceived understanding.
    
    Comparative and attribute claims capture contrasts and observed properties, drawing on fine-grained scientific claim representations~\cite{magnusson2021extractingfinegrainedknowledgegraphs}. A \emph{comparative} claim records the conditions compared, the metric, and the reported difference, such as higher trust in one interface condition than another. An \emph{attribute} claim records a construct, property, and value, such as the mean trust rating in a condition. Each form retains the reported result and its evidentiary context.
    \item \textbf{Entity schema}: Each claim slot is typed by facet, so that a construct is recorded together with the aspect of the study it belongs to. The facets fit the constructs a given literature reports, which makes them the principal point at which the schema is adapted to a domain.
\end{enumerate}

Scope is recorded per \emph{study arm}, an arm being one distinct system-and-population combination. A paper reporting several sub-studies yields one scope description per arm, and each claim references the arm it was extracted from, preserving the system and population on which a finding was measured.

\textbf{Instantiation.} Six facets adapt the components of an HCI study identified by~\cite{li2005intellectual,zhang2009intellectual} and applied to AI-enabled systems by~\cite{rzepka2018user}, namely human, technology, interaction, task context, and outcome. A pilot extraction informed two adaptations to capture finer distinctions in the constructs studied in contemporary human-AI interaction research. First, we distinguish a construct's domain from its role in a finding. A human characteristic such as trust can be an outcome in one claim and a predictor in another. Outcome is therefore represented by a construct's position within a claim, such as the object of an effect relation, while facets describe the constructs themselves. Second, we distinguish system types from system characteristics, and human demographics from human characteristics. For example, a conversational agent names a system type, while its response latency names a characteristic. The resulting facets are human characteristic, human demographic, system characteristic, system type, task context, and interaction.

\begin{figure*}
    \centering
    \includegraphics[width=1\linewidth]{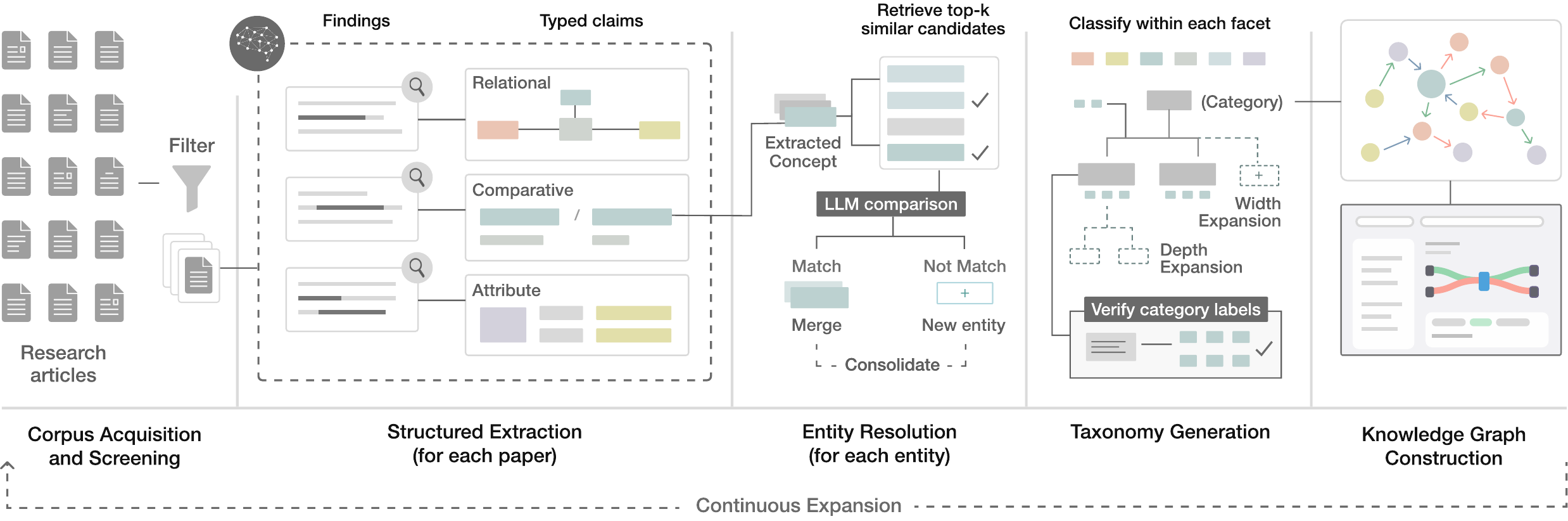}
    \caption{The synthesis pipeline. Screened full-text papers are processed into findings and then relational, comparative, and attribute claims linked to study arms and source evidence. Entity resolution matches extracted terms to canonical entities or admits new entities. Taxonomy generation classifies entities within each facet, expands categories, and verifies their labels. Graph construction integrates these representations for interactive exploration. The pipeline supports subsequent ingestion through continuous expansion.}
    \label{fig:pipeline}
\end{figure*}

Extraction uses OpenAI \texttt{gpt-5.6-luna} at both stages. It yields $308{,}802$ claims over $31{,}741$ papers, spanning $53{,}250$ study arms and $277{,}288$ atomic findings. Study arms exceed papers by two thirds, a single paper frequently reporting on more than one system or population. Claims divide into three statement forms, with $129{,}738$ relational, $95{,}767$ comparative, and $83{,}297$ attribute. Within the relational form, $115{,}689$ claims assert an effect, $9{,}668$ a moderation, and $4{,}380$ a mediation. Figure~\ref{fig:pipeline} illustrates the extraction stages and their integration with entity resolution, taxonomy generation, and graph construction.

\subsection{Entity Resolution}
\label{sec:resolution}

Extracted findings retain the terminology of their source papers to avoid imposing a predefined vocabulary on the constructs they describe. The same construct may therefore appear under different names across studies and research communities. For example, a population may be described as \emph{university students}, \emph{college student}, \emph{tertiary student}, or \emph{higher education learner}. Treating these phrases as distinct concepts would fragment the evidence about that population. Entity resolution builds a shared vocabulary bottom-up by identifying terms that refer to the same construct and consolidating them under a canonical entity. Claims from different papers can then be retrieved together and compared while retaining their original wording and study context.

Each facet is resolved independently. For a facet $f$, let $P_f$ be its pool of extracted phrases and $E_f$ its canonical entity set, which grows as resolution proceeds. Every entity-bearing claim field contributes to $P_f$, including subject, object, moderator, and mediator. Resolution follows the \emph{Selecting} strategy of Wang et al.~\cite{wang2025match}, the strongest of the three strategies evaluated there, in which retrieval proposes candidate entities and a single model call adjudicates among them. Each phrase $p \in P_f$ retrieves a shortlist $S(p) \subseteq E_f$, the $k$ nearest entities above a similarity floor $\tau$. A language model reads $p$ beside $S(p)$ and either selects some $e \in S(p)$, recording $p$ as an alias of $e$, or admits $p$ to $E_f$ as a new entity. Similarity assembles the candidates and the model decides the merge. Resolution decides only whether two phrases name the same entity; whether one entity falls under another is decided over the completed $E_f$ during taxonomy construction (Section~\ref{sec:taxonomy}).

Phrases are resolved in waves ordered by frequency, so that the most common phrasings are canonicalized first and the long tail is judged against an entity set already populated with them. Because a wave is processed against the entity set as it stood at the start of that wave, two phrases naming the same entity can both be admitted as new. Each wave therefore ends with a consolidation pass, in which every entity admitted during the wave is re-evaluated against the others and merged where they match. Consolidation also permits a low $\tau$, since similarity alone does not separate variants of one construct from genuinely different constructs, and raising $\tau$ to exclude the second would also exclude the first.

\textbf{Instantiation.} Phrases are embedded with \texttt{Qwen3-Embedding-8B}, and each $S(p)$ holds the $k = 8$ nearest entities above $\tau = 0.5$. The systems this field studies most are also the most variously named. Papers reporting on a chatbot built over a large language model describe it as an \emph{llm-powered chatbot}, an \emph{llm-based conversational agent}, an \emph{llm dialogue agent}, or an \emph{llm-based copilot}, among $54$ variants that resolution collects into a single entity. Studies of undergraduates are distributed across $16$ names. Resolution yields $126{,}240$ canonical entities carrying $204{,}973$ aliases, and resolves $99.8\%$ of relational claim slots to a canonical entity.

\subsection{Taxonomy Generation}
\label{sec:taxonomy}

Entity resolution establishes equivalence between phrases and produces a flat set of canonical entities for each facet. This set does not capture broader and narrower relationships among distinct constructs. The human demographic facet, for example, contains $7{,}000$ entities, including \emph{undergraduate student} and \emph{higher-education student}. Taxonomy construction organizes these entities into a hierarchy. Both concepts fall under \emph{Enrollment and Degree Status}, within the path from \emph{Educational Status} through \emph{Students by Education Level}. A query at any category retrieves evidence associated with its descendant concepts, enabling readers to move between broad populations and specific groups.

We adapt TaxoAdapt~\cite{kargupta2025taxoadapt}, a corpus-grounded taxonomy construction method reporting strong results against both corpus-driven and LLM-based baselines, to grow each facet's flat entity set $E_f$ into a hierarchy. The root of each hierarchy is the facet itself, such as \emph{human demographic}, and each node beneath it is a category such as \emph{Educational Status}. Entities are classified downward from the root. A node $n$ holds a set of entities $M(n)$, and a language model judges each one against the children $C(n)$ and assigns it to those it belongs under. Entities the model places under no child remain at $n$ as its unassigned set $U(n)$. Construction begins from a shallow seed taxonomy into which entities are classified, then expands along two axes.

\begin{itemize}
  \item \textbf{Depth expansion} subdivides a leaf that has accumulated enough
entities to warrant internal structure, applying when $C(n) = \emptyset$ and $|M(n)| \ge \theta_d$.
  \item \textbf{Width expansion} adds siblings to a node whose existing children
leave too much of its content unaccounted for, applying when $C(n) \neq \emptyset$ and $|U(n)| \ge \theta_w$.
\end{itemize}

In both expansions, a language model assigns each entity a pseudo-label describing it relative to its parent, and these pseudo-labels are clustered into new non-overlapping child nodes, which the model then names. Recursion continues until no node satisfies either condition or the tree reaches a depth ceiling $D$. Floors on child and cluster size bound fragmentation.

We extend TaxoAdapt with a test applied to every generated label. For a node $n$ with candidate label $\ell(n)$ and a sibling $s$, a language model reads $\ell(n)$ without its description, beside a randomized mixture of concepts drawn from $M(n)$ and $M(s)$, and the label is accepted only when it alone identifies which concepts belong to $n$. Labels that fail are regenerated with more specific wording, the target being the established construct name a domain expert would use. TaxoAdapt applies its \emph{sibling coherence} measure after the tree is built, whereas this test determines which labels enter it.

\textbf{Instantiation.} One tree is built per facet. A leaf is subdivided once it holds $\theta_d = 12$ entities, and a node gains siblings once $|U(n)|$ reaches $\theta_w = 10$, with recursion stopping at $D = 4$. A generated node is kept only when $|M(n)| \ge 8$, and a pseudo-label cluster only when it holds at least $3$ concepts. Generation yields six trees totaling $11{,}355$ nodes across five levels, with $6$ roots, $90$ at depth $1$, $805$ at depth $2$, $3{,}940$ at depth $3$, and $6{,}514$ at depth $4$. Label verification holds the resulting categories at the granularity of named constructs. Trust, for instance, is reached through \emph{Motivation, Affect, Attitudes, and Values} and then \emph{Trust, Loyalty, and Commitment}.

\subsection{Knowledge Graph Construction}
\label{sec:graph}

The preceding stages establish complementary representations of the literature. Graph construction integrates them into a unified structure for exploring relationships across studies. Claims link to canonical entities and their positions within facet taxonomies, enabling queries that combine constraints across facets. Relationships are aggregated at two levels. Leaf edges connect canonical entities, while branch edges connect taxonomy categories and aggregate claims associated with their descendants. Both levels draw on the same underlying claims, allowing readers to move from category-level patterns to specific relationships and individual findings, each linked to its study arm and source paper.

\textbf{Instantiation.} Composition yields $548{,}295$ claim-entity links over $308{,}802$ claims. The primary view aggregates at the branch level and the drill-down at the leaf level, since leaf edges are sparse at this corpus size. Median support per edge is $1$ paper at the leaf level and $6$ at the branch level.

\subsection{Continuous Expansion}
\label{sec:maintenance}

Maintaining the Atlas as a resource for research sensemaking requires incorporating new findings as the literature develops. At the rate reported in Section~\ref{sec:intro-gap}, a graph built once omits a substantial body of work within months. The graph is therefore required to be expandable, which the pipeline supports by running periodically over new publications.

Each cycle acquires the publications that have appeared since the last, then screens, retrieves, and extracts from them as before. Resolution runs against the entity set $E_f$ already in the database. A phrase $p$ from a new paper retrieves its shortlist $S(p) \subseteq E_f$ there, and the model records $p$ as an alias of some $e \in S(p)$ or admits it to $E_f$, so evidence about a construct accumulates across cycles. Entities admitted during a cycle pass through the same consolidation used within a wave, then are classified into the existing tree, entering the membership set $M(n)$ of a node already built.

The expansion conditions of Section~\ref{sec:taxonomy} govern structural change. A cycle that raises $|M(n)|$ to $\theta_d$ at a leaf, or $|U(n)|$ to $\theta_w$ at an internal node, triggers expansion on that subtree alone. A node below both thresholds keeps its children. Graph construction projects the extended claim set, and an edge gains support as claims accumulate.

\section{Visualization Interface Design}
\label{sec:design}

The computational pipeline as described in Section~\ref{sec:pipeline} transforms a large, unstructured research corpus into a symbolic representation of findings and their relationships. Making this representation useful for research requires equal attention to how people interpret and explore it. The interface supports this sensemaking process by helping readers examine evidence, follow connections between concepts, and relate individual findings to broader patterns in the literature. Recognizing that researchers may begin with a specific finding to investigate or a broader topic to explore, we developed the interface to support both bottom-up and top-down exploration. The bottom-up path connects individual claims to their evidence, study contexts, and related concepts, while the top-down path moves from an overview of the corpus toward particular concepts and findings. Three design goals, grounded in user experience design and cognitive science, guide how these paths support research sensemaking.

\subsection{Design Goals}
\subsubsection{\textbf{DG 1: Preserve a visible path from every claim to its
  evidence}}  Reading an unfamiliar literature is cumulative: each new finding
  is interpreted in relation to what preceded it and to the evidence on which
  it rests. Interfaces often disrupt this continuity when extracted claims,
  navigation history, and source material are presented separately. Early
  hypertext research identified this loss of orientation and the cognitive
  overhead of reconstructing one's path as central problems in navigating
  linked information~\cite{conklin1987hypertext}. Consistent with the principle
  of recognition rather than recall~\cite{nielsen1994heuristics}, traceability
  should therefore be represented in the interface rather than left for the
  reader to remember.
  
  External representations can also support cognition by
  turning relationships that would otherwise need to be reconstructed mentally
  into objects that can be directly inspected~\cite{kirsh1994epistemic}.
  Educational research similarly shows that explicitly representing and
  attending to source information can improve learners' ability to identify and
  use it when evaluating documents~\cite{britt2002source}. The interface should
  therefore preserve an inspectable chain from each structured claim to its
  study design, sample, population, statistics, source sentence, and paper.
  When a claim enters a conversational synthesis, the selected evidence set and
  claim-level citations should remain visible, allowing the reader to trace an
  answer through the conversation to the evidence from which it was produced.

  \subsubsection{\textbf{DG 2:  Reveal evidence progressively, at the specificity
  the reader requires}} A literature contains more information than can be
  usefully presented in a single view. Working memory is sharply limited, with
  capacity commonly estimated at approximately three to five chunks under
  controlled conditions~\cite{cowan2001magical}, while unnecessary processing
  demands can consume capacity that would otherwise support schema
  formation~\cite{sweller1988cognitive}. The interface should therefore reveal
  complexity progressively rather than presenting every construct, relation,
  and study at once.
  
  Progressive disclosure initially foregrounds the
  information most relevant to the current task while keeping additional
  specificity available on demand~\cite{nielsen2006progressive}; this follows
  the broader HCI principles of matching the system to a user's conceptual
  model~\cite{nielsen1994heuristics} and moving from ``overview first'' to
  filtering and details on demand~\cite{shneiderman1996eyes}. This sequencing
  matters because a coherent model of how a system is organized can support
  faster learning, more accurate retention, and better inference than
  memorizing isolated procedures~\cite{kieras1984mentalmodel}. The interface
  should therefore first orient the reader with a concept's principal incoming
  and outgoing relations, then allow them to reveal individual relations,
  methodological distinctions, claims, and source passages as these become
  relevant. Progressive disclosure does not remove the literature's
  complexity; it lets the reader determine when and at what granularity to
  encounter it.

  \subsubsection{\textbf{DG 3: Scaffold exploration without taking control from
  the reader}} Researchers approaching an unfamiliar area may not yet know the
  constructs or vocabulary needed to formulate a useful query. Exploratory
  search is therefore an iterative process of learning and investigation, not
  simply the retrieval of a previously specified item~\cite{marchionini2006exploratory}.
  Developmental research characterizes scaffolding as structuring a task,
  directing attention toward consequential features, and reducing unnecessary
  degrees of freedom while preserving the learner's active role
  ~\cite{wood1976tutoring}.
  
  Research on learning technologies similarly
  translates scaffolding into interface mechanisms that make processes visible,
  organize complex tasks, and provide guidance at points where learners need
  it~\cite{quintana2004scaffolding}. The interface should accordingly permit
  entry without a query, expose the field's high-level structure, indicate
  evidence-rich paths, and allow readers to descend toward particular
  relationships through reversible expansion and filtering. These supports
  should remain optional: readers must be able to backtrack, change scope,
  follow an unanticipated adjacency, or move directly to detailed evidence,
  consistent with user control and freedom and flexibility and efficiency of
  use~\cite{nielsen1994heuristics}. Because exploration develops across rather
  than within a single interaction, readers should also be able to retain
  claims across sessions, assemble subsets for synthesis, and revisit the
  evidentiary path through which an interpretation emerged.

\subsection{Two Paths Through the Evidence}

Atlas supports two complementary ways of exploring research. In a
\textbf{bottom-up} path, a reader begins with one claim and expands outward to
understand its concept, paper, and wider evidence base. In a \textbf{top-down}
path, the reader begins with the structure of the field and progressively
narrows it to relevant concepts and claims. Cognitive accounts of sensemaking
similarly describe understanding as movement between evidence and an emerging
frame: evidence changes the frame, while the frame guides which evidence is
sought next~\cite{klein2006dataframe,pirolli2005sensemaking}. The two paths are
therefore connected rather than sequential. Every aggregate view can be traced
to individual claims, and every claim can lead back to the larger research
landscape.

\subsection{Bottom-Up Exploration}

\subsubsection{Claim Card (DG1)}

The claim card is the smallest unit in Atlas. It turns an extracted finding
into a concise and inspectable research insight. External representations can
make relationships explicit that would otherwise need to be reconstructed
mentally, reducing the search and computation required to interpret them
~\cite{larkin1987diagram,kirsh1994epistemic}. Relational claims therefore show
a subject, relationship, and outcome; comparative claims show the conditions
and reported difference; and attribute claims show a property and its value.
Using a distinct layout for each claim type prevents unlike findings from being
forced into the same visual structure.

For example, a reader can identify a reported increase from transparency to
trust, inspect its sample and effect size, and verify the source sentence
without locating and searching through the full paper.

\begin{figure}[htbp]
    \centering
    \includegraphics[width=\linewidth]{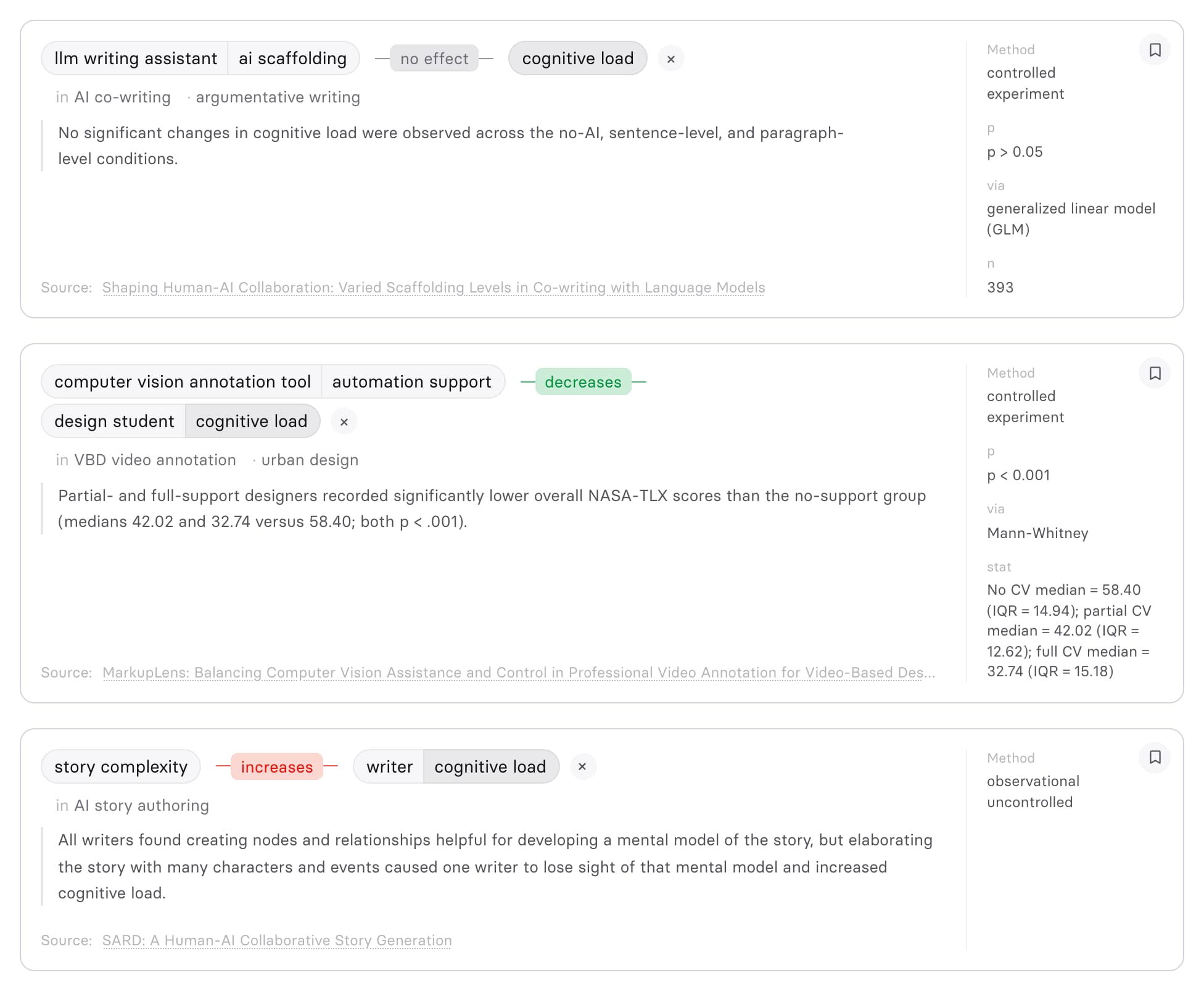}
    \caption{Claim cards showing findings about cognitive load across three study contexts. Each card links the finding to its study context, source passage, and available methodological and statistical details.}
    \label{fig:teaser}
\end{figure}

Each card also reports the study method and, where available, the sample,
population, statistical significance, and effect size. Outcome colour
distinguishes findings that are beneficial, harmful, or neutral, separately
from whether the measured relationship increases or decreases. Readers can
expand the source sentence to verify the extraction or bookmark the claim for
later use. Keeping this information visible follows the principle of
recognition rather than recall~\cite{nielsen1994heuristics} and reflects
research showing that explicit attention to source information supports
readers' ability to identify and evaluate the origins of claims
~\cite{britt2002source}.

\subsubsection{Concept Page (DG2)}

Selecting a concept, such as \emph{trust}, opens a page containing the aggregated evidence connected to that concept. The page reports how many claims and papers
address it and lets the reader switch between what affects the concept and what
the concept affects. This supports the iterative construction and revision of
mental models as readers encounter new evidence
~\cite{kieras1984mentalmodel,klein2006dataframe}.

For example, opening \emph{trust} immediately surfaces its principal reported
causes and effects, after which the reader can move downward to the claims
supporting any selected relationship.

\paragraph{Overview.}
A Sankey diagram summarizes the breadth of the evidence. Band width represents
the number of supporting claims, while colour represents the overall direction
of the reported outcome. Diagrammatic representations can make patterns and
relationships available through perceptual inspection rather than requiring
them to be inferred from sequential text~\cite{larkin1987diagram}. The view
initially emphasizes the strongest relationships so that the diagram remains
readable, while additional relationships remain available through interaction.

For example, the overview can reveal that recommendation explanations are
frequently connected to trust before the reader knows which papers or keywords
would expose that relationship.

\begin{figure}[htbp]
    \centering
    \includegraphics[width=\linewidth]{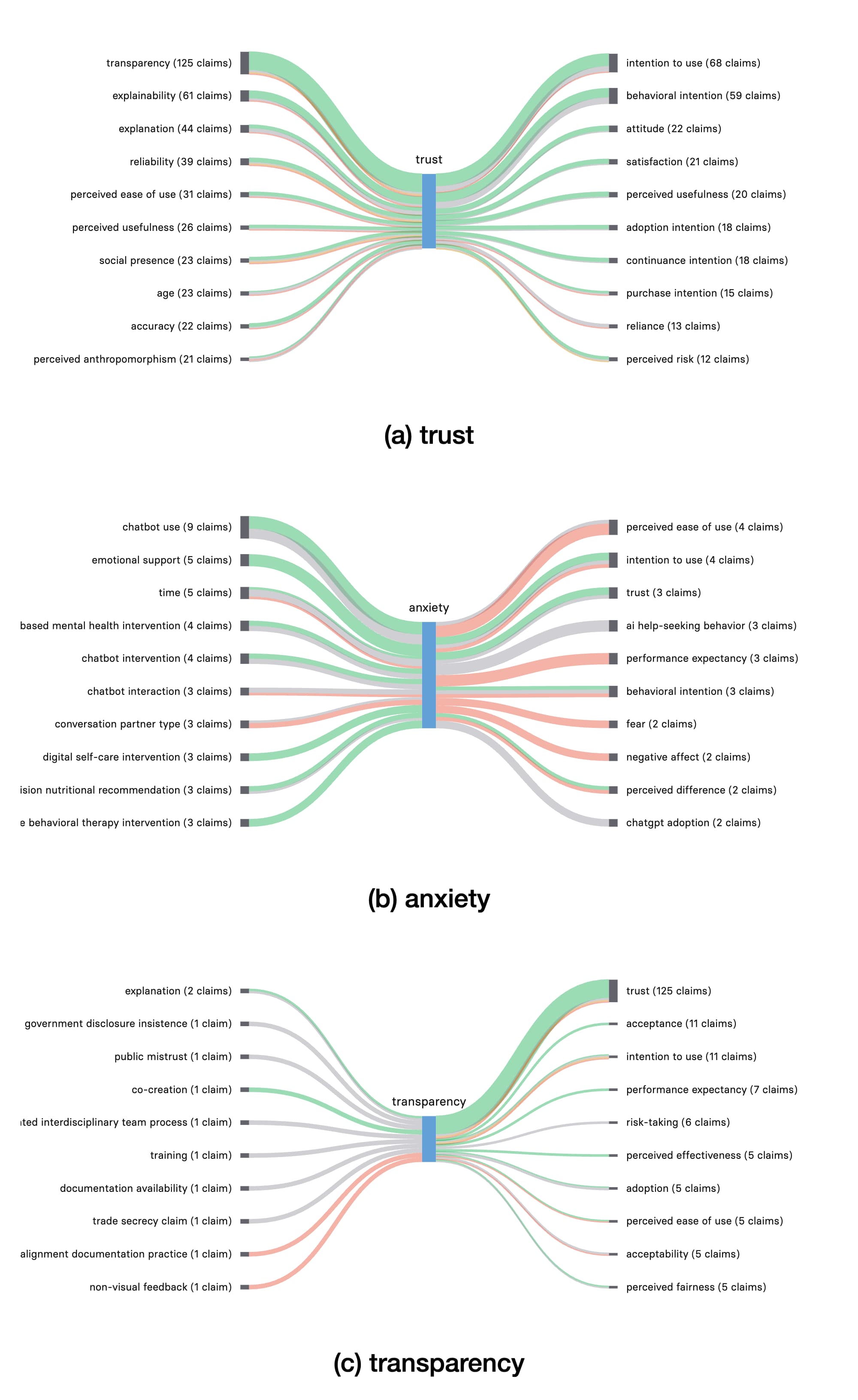}
    \caption{Overview Sankeys. Concept-level relationships around (a) trust, (b) anxiety, and (c) transparency, showing up to ten neighbours on each side. Band width encodes claim count, and colour shows the distribution of reported human outcomes.}
    \label{fig:teaser}
\end{figure}

\paragraph{Detailed evidence.}
The individual claim cards underlying the Sankey appear directly below it.
Readers can filter them by study method, claim type, outcome, publication year,
or population, with counts updating to show how much evidence remains. These
counts provide information scent that helps readers decide which paths are
likely to be useful~\cite{pirolli1999foraging}. The concept page therefore
balances breadth and depth: the Sankey summarizes the research landscape,
while the claim cards expose the findings from which that summary was
produced. This follows the established information-visualization pattern of
providing an overview before filtering and revealing details on demand
~\cite{shneiderman1996eyes}. If a researcher were to select the age--trust relationship here, it might filter their overview to four
claim cards, making their conflicting directions, populations, and methods
directly comparable.

\begin{figure}[htbp]
    \centering
    \includegraphics[width=\linewidth]{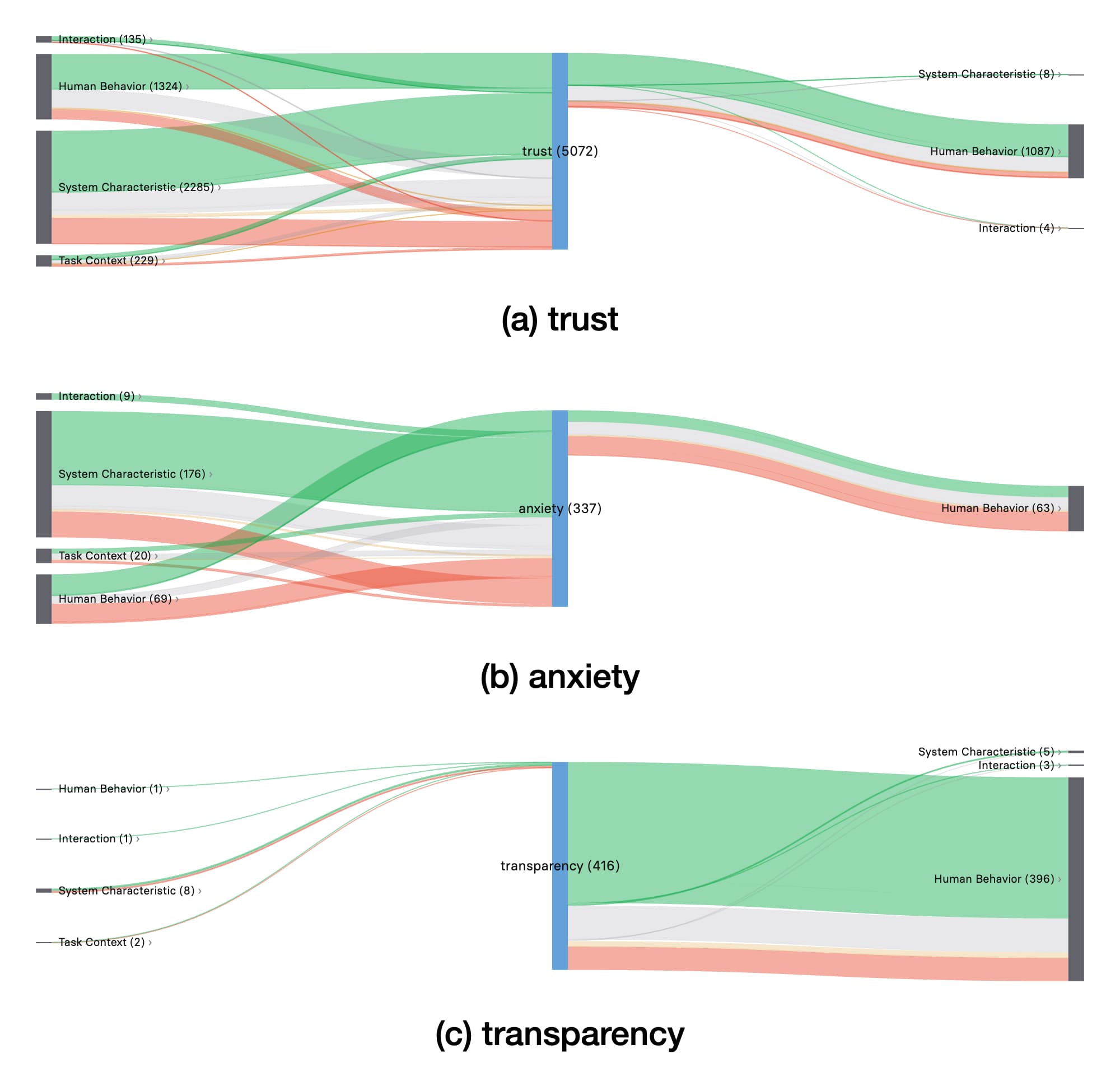}
    \caption{Detailed Sankeys. The same three concepts viewed through taxonomy facets, revealing contrasting distributions of incoming and outgoing evidence. Each category can be expanded independently for closer inspection.}
    \label{fig:teaser}
\end{figure}

\paragraph{Example use cases.}
Figure~\ref{fig:concept-use-cases} illustrates two ways of using the concept
page. A reader interested in \emph{anthropomorphism} can use the overview to
quickly understand the breadth and direction of the surrounding evidence
without opening individual claims. This provides a lightweight overview when
further detail is not needed.

The \emph{trust in AI} example shows a deeper path. The reader first sees that
the concept is connected to 325 claims across 169 papers, together with its
principal causes and effects. Selecting the relation between age and trust then
filters the page to four underlying claims. The claim cards show that the
reported relationship is not uniform: three claims report a decrease in trust,
while one reports an increase, across different populations, contexts, and
study methods. The reader can therefore move from the breadth of a concept to
the specific evidence behind one relationship.

\begin{figure*}[t]
    \centering

    \begin{minipage}[t]{0.48\textwidth}
        \vspace{0pt}
        \centering
        \includegraphics[width=\linewidth]
        {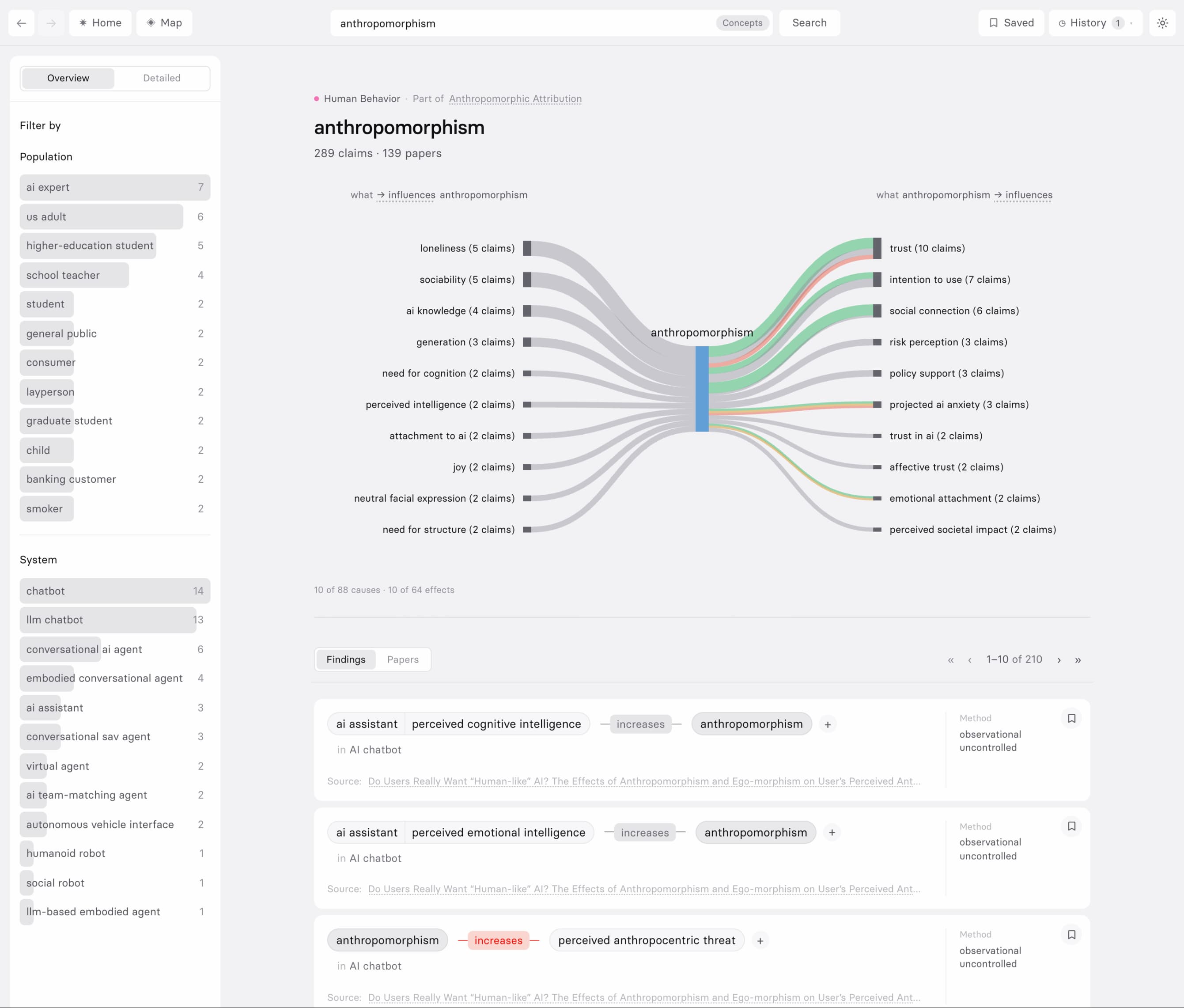}

        \smallskip
        \textbf{(a) Concept View: Anthropomorphism  (Overview)}
    \end{minipage}
    \hfill
    \begin{minipage}[t]{0.48\textwidth}
        \vspace{0pt}
        \centering
        \includegraphics[width=\linewidth]
        {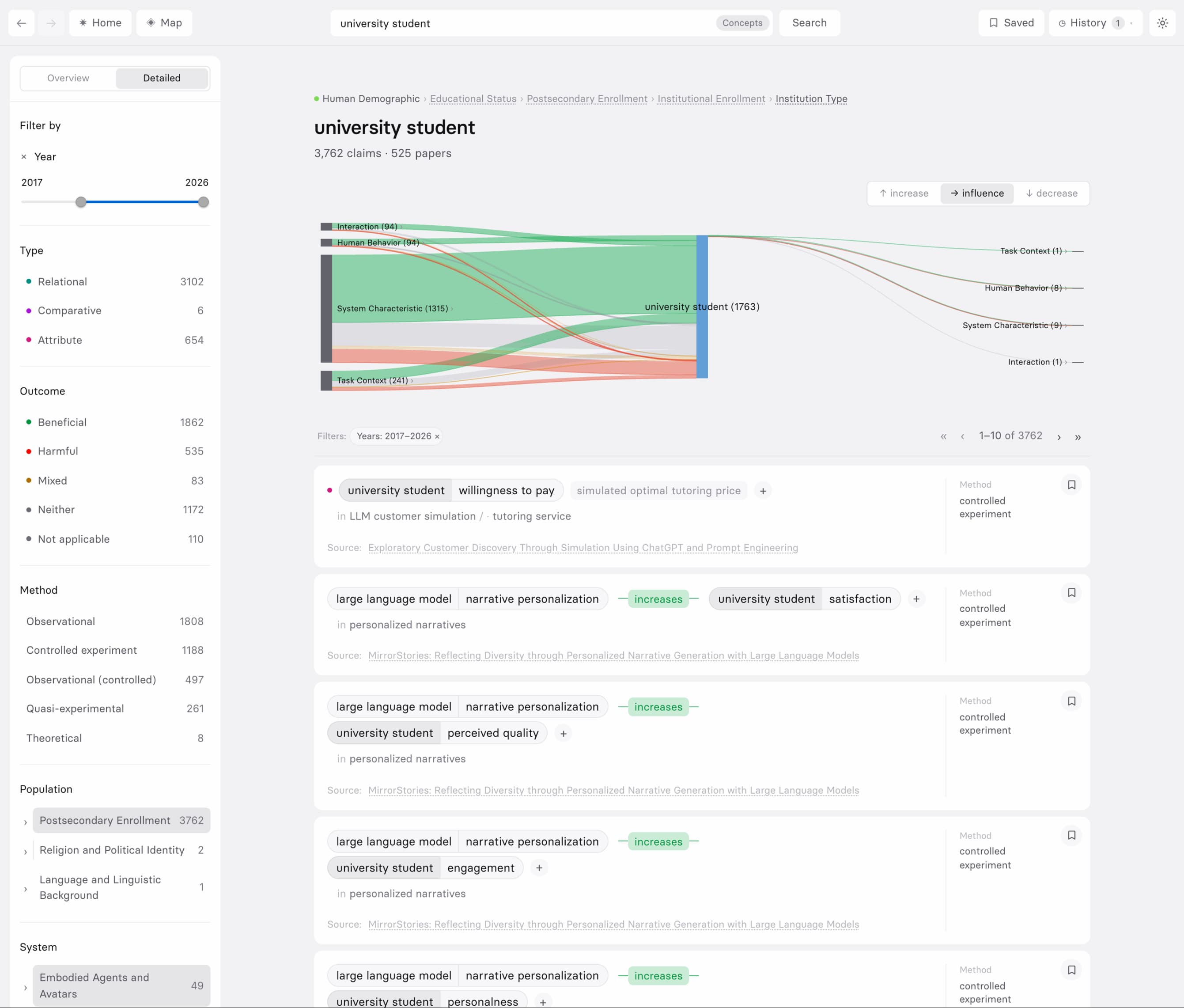}

        \smallskip
        \textbf{(b) Concept View: University Student  (Detailed)}
    \end{minipage}

    \caption{\textbf{Two concept-page use cases.}
\textbf{(a)} The anthropomorphism overview shows incoming and outgoing
relationships alongside supporting claims.
\textbf{(b)} The university student detailed view groups connections
by facet and presents individual claims with filters for narrowing
the evidence.}
    \label{fig:concept-use-cases}
\end{figure*}

\subsubsection{Paper View (DG1)}

Selecting a claim's source opens a structured view of the original paper. The
page retains the paper's bibliographic information, abstract, and source link,
while separately highlighting its core research questions, study designs,
samples, and key findings. This structure helps the reader form a coherent
model of the study before engaging with its full detail; coherent mental models
have been shown to support learning, retention, and inference
~\cite{kieras1984mentalmodel}.
A researcher, after encountering an unfamiliar claim in the graph, a reader can
open its paper view to inspect the study's research question, sample, method,
and other extracted findings in context.

The paper view uses progressive disclosure by presenting the information most
important for initial understanding while keeping the complete source
available~\cite{nielsen2006progressive}. This reduces extraneous processing
that could otherwise compete with the reader's effort to understand the
study~\cite{sweller1988cognitive,cowan2001magical}, without treating the
structured view as a replacement for reading the original paper.

\subsection{Top-Down Exploration}

Readers who do not yet have a specific concept in mind can begin with the
corpus views. Exploratory search often begins without a precise query and
develops as the reader learns new structures and terminology
~\cite{marchionini2006exploratory}. Information-foraging research likewise
describes search as movement between information patches, guided by cues about
the likely value and cost of following each path~\cite{pirolli1999foraging}.

Atlas supports this process through a toggle between the \emph{Graph} and
\emph{Flow} views. The Graph view supports open discovery, while the Flow view
provides a more guided route into the evidence. Allowing readers to choose
between these representations accommodates different levels of expertise and
exploratory intent, consistent with flexibility, efficiency, and user control
~\cite{nielsen1994heuristics}.

\subsubsection{Graph View (DG3)}

The Graph view presents the corpus as a three-dimensional research landscape.
It combines the taxonomy with the claim network: taxonomy links preserve the
field's hierarchy, while claim links bring concepts together when the
literature reports relationships between them. Height represents taxonomy
depth, with broad facets above more specific concepts.

The purpose of this view is open exploration rather than precise comparison.
Readers can scan the field for dense regions, isolated areas, and unexpected
connections without first knowing what to search for. Selecting a node reveals
its immediate relationships and opens its concept page. By externalizing the
field's structure, the Graph view makes relationships available as part of the
reader's cognitive environment rather than requiring them to be held and
combined internally~\cite{hollan2000distributed}. It thereby supports the
discovery of concepts that a query-based interface would require the reader to
name in advance~\cite{marchionini2006exploratory}.
For example, following an unfamiliar branch from \emph{recommendation
explanations} can reveal adjacent concepts that the reader would not have known
to include in an initial keyword query.

\paragraph{From the graph to detailed evidence.}
In the Graph view \ref{fig:concept-use-cases}, selecting \emph{Recommendation Explanations} reveals a branch containing 13 concepts and 62 claims. The side panel presents the claims connected to that branch, allowing the reader to inspect their direction, context, and sources. Selecting a source then opens the detailed paper view, which presents the paper's abstract, research questions, study design, and findings. The reader can therefore move from a broad and potentially unfamiliar region of the literature to the evidence reported by an individual study. This extends the ``overview first, zoom and filter, then details on demand'' pattern while preserving a traceable path back to the source~\cite{shneiderman1996eyes}.

\begin{figure*}

    \begin{minipage}[t]{0.48\textwidth}
        \vspace{0pt}
        \centering
        \includegraphics[width=\linewidth]
        {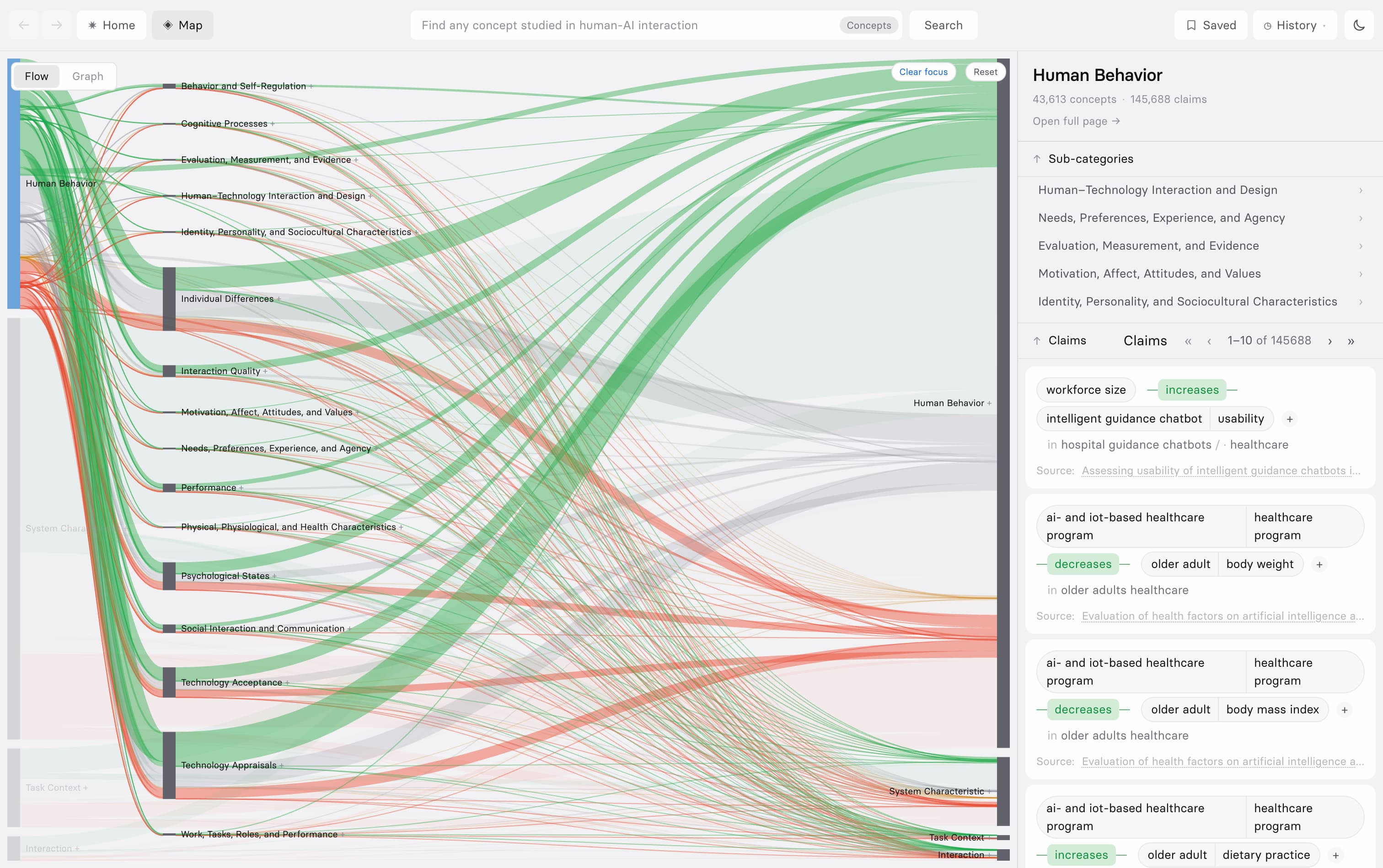}

        \smallskip
        \textbf{(a) Flow View}
    \end{minipage}
    \hfill
    \begin{minipage}[t]{0.48\textwidth}
        \vspace{0pt}
        \centering
        \includegraphics[width=\linewidth]
        {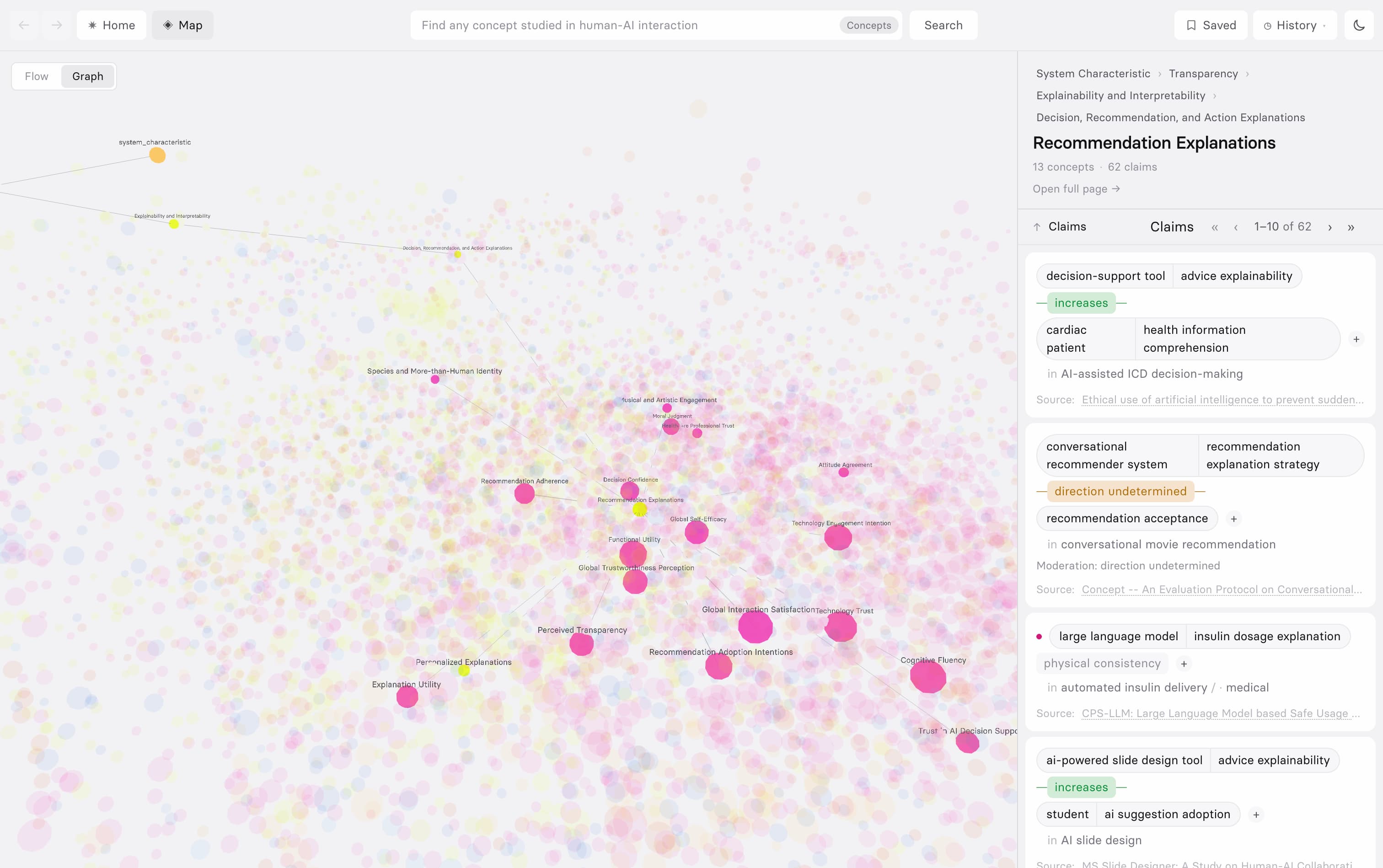}

        \smallskip
        \textbf{(b) Graph View}
    \end{minipage}

    \caption{The \textit{Flow} and \textit{Graph} in the Top-down view.     \textbf{(a)} The \textit{Flow} view shows relationships across research categories, with \emph{Human Behavior} selected to reveal its subcategories and associated claims.
    \textbf{(b)} The \textit{Graph} view presents the broader research landscape, with \emph{Recommendation Explanations} selected to inspect its connections and underlying claims.}
    \label{fig:concept-use-cases}
\end{figure*}

\subsubsection{Flow View (DG3)}

The Flow view offers a more structured top-down path. It begins with broad
research categories, such as system characteristics, human behavior, task
context, and interaction. Bands connect reported causes and effects; their
width represents the number of claims and their colour represents the overall
outcome.

Readers can expand any category into more specific concepts while leaving the
rest of the field at a broader level. Selecting a band filters the interface to
the claims behind that relationship. A reader can therefore move from a
field-level pattern, to a specific relationship, and finally to individual
claim cards without formulating a query. This progressive expansion functions
as a scaffold: it reduces the complexity of the initial task while allowing
the reader to take greater control as their understanding develops
~\cite{wood1976tutoring,quintana2004scaffolding}. It also applies the
``overview first, zoom and filter, then details on demand'' principle to the
literature itself~\cite{shneiderman1996eyes,nielsen2006progressive}.

For example, a reader can expand \emph{system characteristics} into increasingly
specific constructs and then select one band to inspect the individual claims
behind the field-level pattern.

\subsubsection{Category Page (DG2)}

Selecting a taxonomy category opens a page that combines all concepts and
claims within that branch. The page shows the category's children, the concepts
beneath them, and the evidence attached to the resulting set.

This helps readers test and revise whether their initial framing is too broad
or too narrow~\cite{klein2006dataframe}. For example, a search for
\emph{students} can be widened to educational status or narrowed to
undergraduate, graduate, or degree-progression populations. These distinctions
change which findings apply and prevent different populations from being
treated as a single body of evidence. By making the underlying conceptual
structure visible, the category page also supports the development of a more
accurate mental model of the literature~\cite{kieras1984mentalmodel}.
A reader who uses the \textit{category page} feature can broaden \emph{university students} to educational
status or narrow it to undergraduate populations and immediately observe how
the applicable evidence changes.

\subsection{Collection and Synthesis (DG3)}

Both exploratory paths lead to a persistent collection. Readers can bookmark
claims from any surface and retain them across sessions. The collection appears
beside a conversational synthesis pane, allowing each cited claim to remain
visible while the synthesis is read. The collection therefore acts as an
external memory: earlier findings remain available for inspection rather than
having to be recalled or reconstructed
~\cite{hollan2000distributed,kirsh1994epistemic,nielsen1994heuristics}. 

\paragraph{Scoping a question.}
Readers choose which saved claims are relevant before beginning a conversation.
The selected set is then held fixed, ensuring that subsequent answers refer to
a stable body of evidence. Claims can be added as the exploration develops,
and the same collection can support several questions through different
subsets. This turns exploration into a cumulative sensemaking process while
preserving a visible path from each synthesized statement to the individual
findings on which it rests~\cite{pirolli2005sensemaking}

\paragraph{Traceability.}
The synthesis is designed so that readers can trace each conclusion back to its evidence. The agent uses only the papers linked to the selected claims and returns a direct answer, key points, disagreements, and numbered citations. Each point identifies its supporting claims, allowing readers to move from the response to the claim, source sentence, and original paper. Keeping this evidence visible reduces the need to remember or reconstruct the reasoning path, consistent with research on external representations and distributed cognition~\cite{larkin1987diagram,hollan2000distributed}. Explicit source information also helps readers evaluate how evidence is being used~\cite{britt2002source}.

Two controls keep the synthesis bounded and auditable. First, retrieval is limited to the papers behind the selected claims, so the response does not silently introduce evidence from elsewhere in the corpus. Second, quantitative summaries are computed directly from the database rather than estimated by the model. Counts, outcome directions, study designs, and paper contributions therefore remain tied to the underlying data.

\begin{figure}[htbp]
    \centering
    \includegraphics[width=\linewidth]{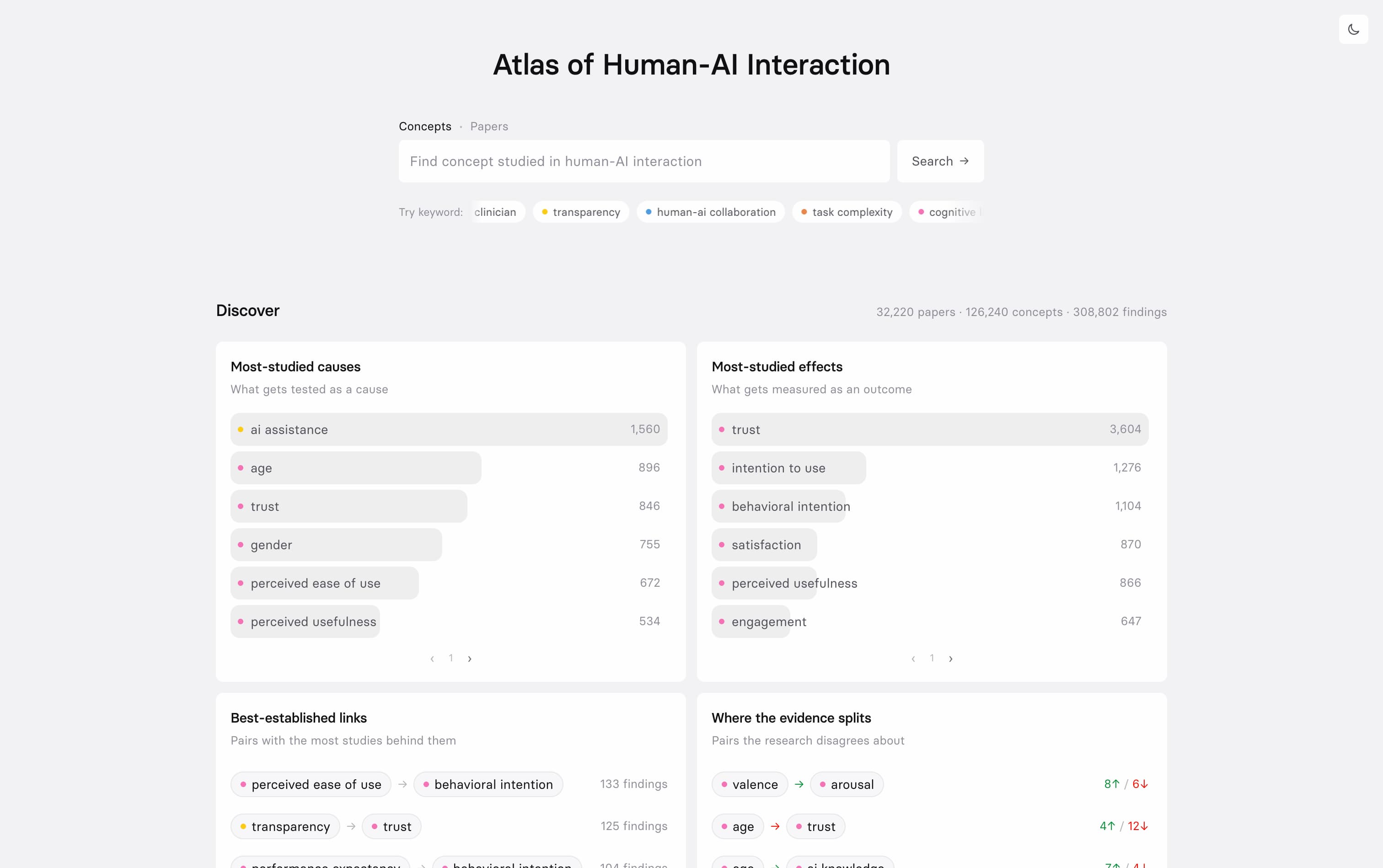}
    \caption{\textbf{Discovery and navigation in Atlas.} Persistent tabs preserve
    the reader's exploratory branches, filters, and position while consistent
    claim representations connect overview and evidence views.}
    \label{fig:discovery-navigation}
\end{figure}

\subsection{Discovery and Navigation}

Atlas supports discovery without requiring readers to abandon or reconstruct
their current line of inquiry. Opening a concept, paper, or evidence flow
creates a new tab while leaving the originating view available. Each tab
preserves its filters and scroll position, allowing readers to compare branches,
return to an earlier interpretation, or close an unproductive direction. This
supports user control and freedom while favoring recognition over recall: the
reader's navigation state remains visible rather than having to be remembered
and reconstructed~\cite{nielsen1994heuristics}. It also reduces the
disorientation and cognitive overhead associated with navigating deep
hypertext structures~\cite{conklin1987hypertext} and supports the revisitation
that commonly occurs during information seeking~\cite{tauscher1997people}.For instance, a reader can leave a filtered trust page open, investigate one
source paper in another tab, and return to the original comparison with its
state intact.

The same interaction and visual conventions recur across the landing page,
Graph and Flow views, concept pages, and claim cards. For example, selecting a
relationship always exposes its supporting claims using the same card layout,
outcome colours, and source controls. This follows the consistency and
standards heuristic: readers can learn Atlas's visual language once and direct
their attention toward interpreting the evidence rather than relearning the
interface in each view~\cite{nielsen1994heuristics}. Recent HCI systems similarly
show how structured representations can scaffold scholarly navigation:
Synergi organizes papers into revisable research threads that help scholars
broaden and refine an inquiry~\cite{kang2023synergi}, while DimInd guides
readers across papers, comparison tables, taxonomies, and syntheses with lower
reported effort than a ChatGPT-assisted baseline~\cite{fok2025facetstaxonomiessynthesesnavigating}.

Discovery remains possible at multiple levels of specificity. Readers may
search for a known concept or paper, begin from a live claim on the landing
page, or browse the Graph and Flow views when they do not yet know the relevant
terminology. Because each route ultimately uses the same concepts, claims, and
paper views, readers can move between open-ended discovery and targeted
inspection without changing interaction models. For example, a reader who
encounters \emph{recommendation explanations} on the landing page can follow
the concept into the graph, inspect its connected claims, and open the source
paper without first formulating a sequence of separate searches.
\subsection{Bookmarks}

\begin{figure}[htbp]
    \centering
    \includegraphics[width=\linewidth]{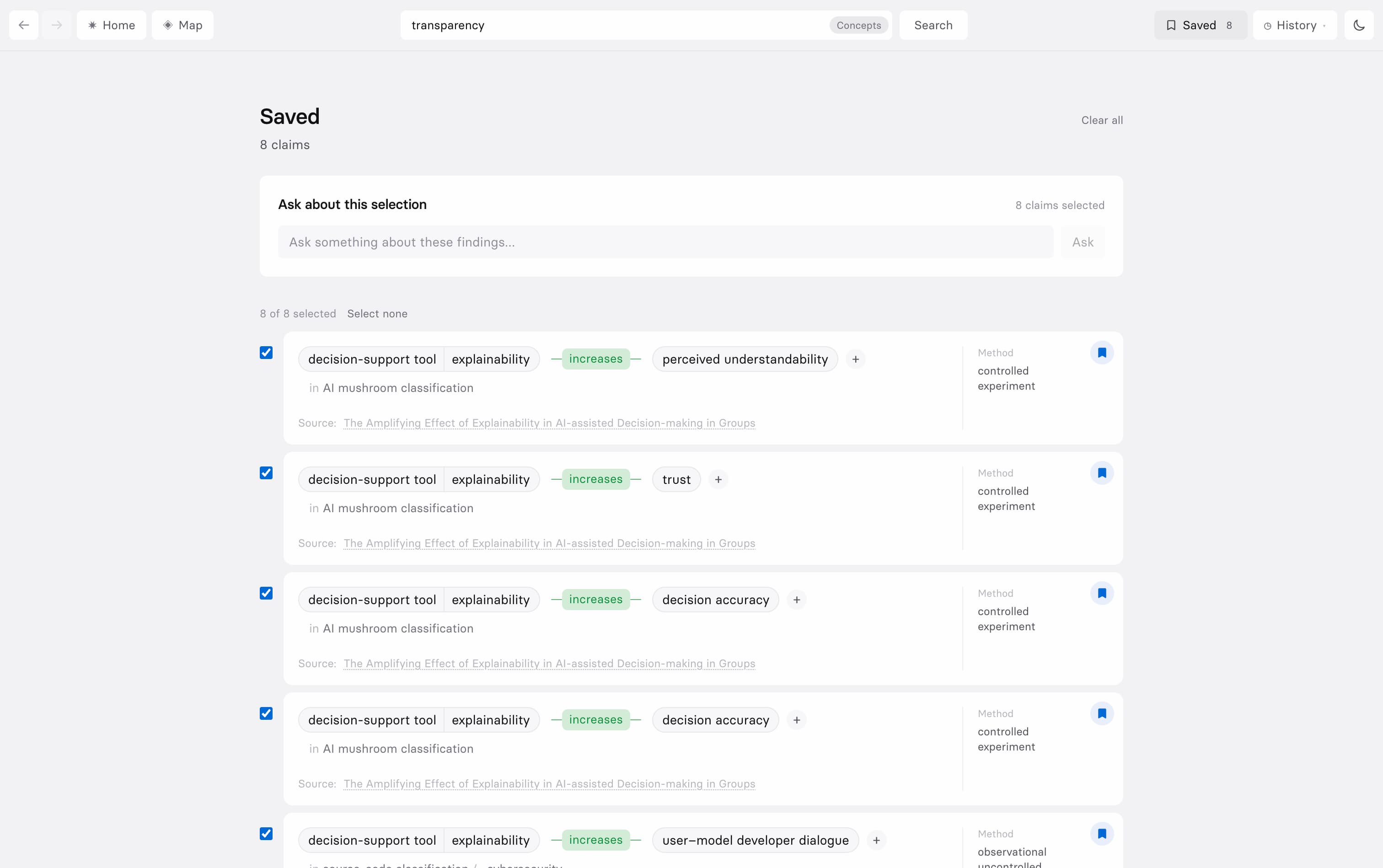}
    \caption{\textbf{Saved claims.} Readers collect findings encountered across
    Atlas into persistent folders that can be revisited, compared, and selected
    as evidence for conversational synthesis.}
    \label{fig:saved-claims}
\end{figure}
Readers can bookmark a claim from any Atlas view and organize saved claims into
persistent collections. We designed these collections around the familiar
folder metaphor used by desktop file-management systems such as Apple's Finder:
an item is placed in a named location, remains there across sessions, and can
later be reopened or reorganized. Reusing this familiar real-world model makes
the consequences of saving predictable, consistent with the heuristic of
matching the system to the real world~\cite{nielsen1994heuristics}. Rather than
requiring readers to learn an Atlas-specific archival mechanism, collections
behave like recognizable containers for research material.

Saving also serves a cognitive function. Research on personal information
management characterizes bookmarking as more than recording an address: people
construct personal information spaces in which selected resources can be
organized and recovered for later work~\cite{abrams1998information}.
Observational studies of ``keeping found things found'' likewise show that
people preserve useful Web information because anticipated reuse cannot be
supported reliably through memory or repeated search alone
~\cite{jones2001keeping,jones2002once}. In Atlas, the collection externalizes
the reader's evolving judgment about which findings matter, reducing the need
to remember both the claims and the paths through which they were discovered
~\cite{hollan2000distributed,kirsh1994epistemic}.

The collection also bridges discovery and synthesis. Readers can revisit saved
claims, place related findings together, and select a subset as the bounded
evidence for a conversation. GPT-5.6 Terra then synthesizes only the papers
associated with that selected claim set, identifying patterns, disagreements,
and qualifications while retaining links to the underlying evidence. This
reduces the demands of simultaneously holding, comparing, and expressing
findings across many studies, tasks that can become difficult as the evidence
set grows or when readers face cognitive or linguistic barriers. The generated
account remains an inspectable starting point rather than a replacement for the
reader's judgment: each statement can be traced to its supporting claims,
source passages, and papers. This division of labor makes synthesis more
accessible while preserving reader control and evidentiary accountability.

Importantly, readers construct the evidence set themselves rather than
passively receiving a preassembled synthesis. This reflects constructionist
accounts of learning, in which people develop understanding by actively
building personally meaningful and inspectable knowledge artifacts
~\cite{papert1980mindstorms,papert1991situating}. The collection becomes such
an artifact: its organization records how the reader has framed the topic and
can be revised as that understanding develops. Related work on the IKEA effect
shows that successful self-assembly can also increase people's valuation of
what they construct~\cite{norton2012ikea}; although this does not itself
demonstrate greater learning, it suggests why assembling a research collection
may strengthen readers' investment in the resulting inquiry.

For example, a reader investigating trust can save four conflicting claims
about age and trust, compare their populations and methods, and ask Atlas to
synthesize only that selected evidence. The reader can then revise the
collection, regenerate the account, and move between the synthesis and the
individual findings from which it was constructed.

\section{Use-Case Scenarios}
\label{sec:use-cases}
Three illustrative journeys show how users can use Atlas to assess evidence coverage, compare outcomes, and refine practical questions. 

\subsection{Informing Policy on AI Companions for Teenagers}
\label{sec:use-case-policy}

\emph{RQ1: What benefits and harms of AI companions are supported by evidence relevant to teenagers?}

A policymaker uses Atlas's \emph{Concept Search} to open \emph{AI companion chatbot}, which contains 139 claims from 29 papers. Switching to \emph{Detailed} view and applying the \emph{Outcome} filters reveals 28 beneficial claims from 13 papers (44.8\%) and 66 harmful claims from 17 papers (58.6\%). Atlas makes both the balance of outcomes and the concentration of multiple claims within papers visible, prompting the policymaker to investigate which populations and experiences underpin these counts.

The policymaker then uses Atlas's \emph{Concept Overview} and selects the \emph{adolescent} population filter. Only one claim remains in the selected direction, concerning content filtering and disengagement. Using \emph{Show evidence} and opening its source in Atlas's \emph{Paper View} reveals a qualitative analysis of 318 Reddit posts, with reports spanning emotional support, sleep, schoolwork, and offline relationships.

To examine benefits, the policymaker uses Atlas's \emph{Paper Search} to retrieve loneliness research and checks its adult study population. Saving these findings alongside the adolescent claims on Atlas clarifies what the aggregate percentages cannot establish: whether short-term relief translates into healthy everyday use by teenagers. The journey produces a focused policy agenda around adolescent-specific evidence, sustained outcomes, and the conditions under which benefits and harms coexist.

\subsection{Scoping a Review of AI in Education for Low-Income Children}
\label{sec:use-case-education}

\emph{RQ2: Which AI supports improve education for low-income children, and how directly does the evidence address this population?}

A researcher uses Atlas's \emph{Concept Search} to open \emph{AI tutor}, then switches to \emph{Detailed} view. Atlas lists 66 claims from 21 papers. The \emph{Beneficial} filter returns 46 claims from 16 papers (76.2\%), while \emph{Harmful} returns nine claims from six papers (28.6\%). This gives the researcher an initial picture of promising outcomes and reported problems, before assessing relevance to economically disadvantaged children.

Searching separately for \emph{low-income student} on Atlas retrieves just one claim from one paper. Opening the evidence in \emph{Paper View} identifies an audit of synthetic student profiles, establishing a distinction between recommendation bias and measured learning. The researcher uses Atlas's \emph{Household Socioeconomic Status} category, containing 16 concepts and 41 claims, to broaden the search vocabulary. This sparse concept result flags a retrieval gap without implying that only one relevant study exists.

A separate \emph{Paper Search} on Atlas retrieves \emph{Tutor CoPilot}: exit-ticket passing increased from 62\% to 66\%, and 67\% of the student sample was economically disadvantaged. The researcher uses Atlas to save learning, tutor-effectiveness, and grade-appropriateness claims for a review matrix. The journey turns an aggregate pattern into a sharper review question about population coverage, educator support, and whether reported gains extend to low-income children.

\subsection{Selecting AI Features to Improve Workplace Productivity}
\label{sec:use-case-productivity}

\emph{RQ3: Which AI assistant features should a company test to improve performance while preserving users' understanding?}

A product team uses Atlas's \emph{Map} in \emph{Flow} mode to explore \emph{Human Behavior} and \emph{Performance}, then searches for \emph{productivity}. In \emph{Detailed} view, Atlas lists 154 claims from 110 papers across settings. Applying the outcome filters retrieves 105 beneficial claims from 82 papers (74.5\%) and 12 harmful claims from 11 papers (10.0\%). The team uses this broad pattern to identify opportunities while inspecting the smaller body of adverse findings.

To narrow the inquiry to coding-assistant features, the team uses Atlas's \emph{Paper Search} to open Google's IDE study and follows \emph{code-diff rendering} into \emph{Detailed} view. Expanded evidence shows that clearer diffs reduced review time by 7\% and increased acceptance by 2.2\% in the same experiment. The team also retrieves and saves a comprehension finding from a student study, adding an outcome that acceptance metrics alone do not capture. The journey helps the team distinguish a promising feature from assumptions that still require evaluation with professional developers.

\section{Analysis of Empirical Human-AI Interaction Research}
\label{sec:analysis}

This section provides a descriptive overview of the collected literature
using the structure produced by the synthesis pipeline. We examine
frequently reported constructs, their organization within the facet
taxonomies, and the relationships connecting them across studies.
The analysis illustrates how the extracted representation supports
examination of corpus-level patterns and offers starting points for
closer exploration of individual findings and their study contexts.

\subsection{Common Constructs and Relationships}

The corpus contains $308{,}802$ claims from $31{,}741$ paper records, organized around $126{,}240$ canonical entities. Claims comprise $129{,}738$ relational, $95{,}767$ comparative, and $83{,}297$ attribute statements. Among papers contributing claims, annual publication counts rise from $1{,}154$ in 2021 to $8{,}743$ in 2025. The $9{,}433$ records dated 2026 reflect an incomplete publication year and are excluded from this comparison.

\emph{AI assistance} is the most frequent subject ($1{,}416$ claims), followed by \emph{age} ($864$) and \emph{trust} ($747$). The most frequent objects are \emph{trust} ($3{,}234$), \emph{intention to use} ($1{,}175$), and \emph{behavioral intention} ($909$). Trust consequently provides an entry point both for examining responses to systems and for following relationships with subsequent attitudes and intentions.

Across all claim forms, common population entities include \emph{university student} ($3{,}791$ claims), \emph{student} ($3{,}443$), and \emph{consumer} ($2{,}679$). Common system entities include \emph{LLM chatbot} ($10{,}593$), \emph{large language model} ($10{,}019$), and \emph{chatbot} ($8{,}828$). These counts include direct entity links and associated population or system tags. Related labels remain separate canonical entities at this level; category views provide broader groupings.

Table~\ref{tab:common-relations} reports the five most frequent effect relationships. Transparency and trust appear alongside relationships among ease of use, perceived usefulness, and intention, while AI assistance and diagnostic accuracy connect system use to measured performance. Claim counts indicate how often a relationship is represented, with each paper potentially contributing several findings.

\begin{table}[t]
\centering
\small
\caption{The five most frequent canonical subject--object pairs among relational effect claims. Counts include all reported directions. Paper counts refer to distinct publication records within each row.}
\label{tab:common-relations}
\begin{tabularx}{\linewidth}{@{}Xrr@{}}
\toprule
Reported relationship & Claims & Papers \\
\midrule
Transparency $\rightarrow$ trust & 121 & 116 \\
Perceived ease of use $\rightarrow$ behavioral intention & 106 & 98 \\
Perceived ease of use $\rightarrow$ perceived usefulness & 101 & 97 \\
Perceived ease of use $\rightarrow$ intention to use & 88 & 72 \\
AI assistance $\rightarrow$ diagnostic accuracy & 83 & 55 \\
\bottomrule
\end{tabularx}
\end{table}

\subsection{Taxonomy Structure and Connections Across Facets}

The six facet taxonomies organize $126{,}240$ canonical entities into $11{,}355$ nodes, including six roots and $90$ categories directly below them. The hierarchy extends to depth four, with $6{,}514$ nodes at that depth. Table~\ref{tab:facet-taxonomies} shows how this structure is distributed. System characteristics contain the largest vocabulary ($47{,}107$ entities), while human characteristics have the most taxonomy nodes ($3{,}440$). These counts describe the organization produced by the pipeline; category size and subdivision reflect both the extracted constructs and the taxonomy generation procedure.

\begin{table}[t]
\centering
\small
\caption{The six facet taxonomies. Node counts include the facet root; top-level categories are its immediate children.}
\label{tab:facet-taxonomies}
\begin{tabularx}{\linewidth}{@{}Xrrr@{}}
\toprule
Facet & Entities & Nodes & Top-level \\
\midrule
Human characteristic & $43{,}613$ & $3{,}440$ & 16 \\
Human demographic & $7{,}000$ & $1{,}214$ & 6 \\
Interaction & $5{,}709$ & $1{,}156$ & 18 \\
System characteristic & $47{,}107$ & $2{,}371$ & 17 \\
System type & $11{,}563$ & $1{,}819$ & 14 \\
Task context & $11{,}248$ & $1{,}355$ & 19 \\
\bottomrule
\end{tabularx}
\end{table}

The facets provide complementary ways to locate evidence. Human and system characteristics describe constructs within findings, while population and system-type tags locate those findings in the groups and technologies studied. Task context and interaction describe further aspects of the activity. Across all claim forms, $79{,}001$ claims ($25.6\%$) link to canonical entities in at least three facets, counting both direct entity links and associated tags. A claim can consequently participate in several views of the corpus, such as a view organized around a human outcome and another restricted to a particular system or population.

\subsection{Research Directions, Relation Types, and Methods by Area}

At the facet level, the largest subject--object flow connects system characteristics to human characteristics, accounting for $52{,}139$ of the $115{,}644$ effect claims with both endpoints resolved ($45.1\%$). Relationships between human characteristics account for $29{,}292$ claims ($25.3\%$), followed by relationships between system characteristics ($10{,}756$; $9.3\%$) and from task context to human characteristics ($9{,}051$; $7.8\%$). The represented research thus includes both evaluations of system properties and relationships among human attributes, attitudes, and behaviors.

To examine how these patterns vary by area, we group relational claims by the depth-one taxonomy category of their human-characteristic object. This rule assigns $99{,}504$ of the $129{,}730$ relational claims with a recognized subtype to an area. Eight further relational records have no subtype and are excluded. Each included claim belongs to one area; a paper may contribute to several. Table~\ref{tab:area-profiles} reports the five largest areas by claim count. Relation types and method profiles use all included relational claims in each area. Subject-facet shares use effect claims with a resolved subject. Study-method profiles are summarized through the extracted evidentiary-basis categories.

The areas differ in what occupies the subject position. In \emph{Performance}, $67.6\%$ of resolved effect subjects are system characteristics, as are $64.8\%$ in \emph{Psychological States}. In \emph{Technology Acceptance}, human characteristics account for $54.4\%$, compared with $33.7\%$ for system characteristics. These profiles distinguish areas organized primarily around system-to-human relationships from an area in which relationships among human constructs are prominent.

\begin{table*}[t]
\centering
\small
\caption{Profiles of the five largest outcome areas. All percentages are within area. In (a), the leading subject facet is measured among effect claims with resolved subjects; relation-type shares use all classified relational claims. In (b), the same relational claims are grouped by extracted evidentiary basis. HC denotes human characteristic; SC denotes system characteristic. Rounding may prevent percentages from summing to 100.}
\label{tab:area-profiles}
\textbf{(a) Subject direction and relation types}\smallskip

\begin{tabularx}{\textwidth}{@{}Xrrrrr@{}}
\toprule
Outcome area & Claims & Leading subject & Effect & Moderation & Mediation \\
\midrule
Technology Appraisals {\footnotesize (e.g., trust, perceived usefulness)} & $22{,}636$ & SC (55.4\%) & 88.0\% & 9.2\% & 2.8\% \\
Technology Acceptance {\footnotesize (e.g., intention to use, AI reliance)} & $21{,}191$ & HC (54.4\%) & 83.0\% & 8.8\% & 8.2\% \\
Performance {\footnotesize (e.g., diagnostic accuracy, task completion time)} & $19{,}078$ & SC (67.6\%) & 90.5\% & 7.4\% & 2.1\% \\
Psychological States {\footnotesize (e.g., cognitive load, anxiety)} & $12{,}837$ & SC (64.8\%) & 89.9\% & 7.1\% & 3.0\% \\
Interaction Quality {\footnotesize (e.g., user satisfaction, engagement)} & $6{,}955$ & SC (62.3\%) & 87.5\% & 7.2\% & 5.2\% \\
\bottomrule
\end{tabularx}

\medskip
\textbf{(b) Study-method profiles}\smallskip

\begin{tabularx}{\textwidth}{@{}Xrrrrr@{}}
\toprule
Outcome area & Controlled exp. & Quasi-exp. & Controlled obs. & Uncontrolled obs. & Theoretical \\
\midrule
Technology Appraisals {\footnotesize (e.g., trust, perceived usefulness)} & 43.8\% & 4.3\% & 10.1\% & 41.2\% & 0.6\% \\
Technology Acceptance {\footnotesize (e.g., intention to use, AI reliance)} & 29.7\% & 4.0\% & 17.7\% & 48.3\% & 0.3\% \\
Performance {\footnotesize (e.g., diagnostic accuracy, task completion time)} & 49.4\% & 11.2\% & 6.6\% & 31.8\% & 0.8\% \\
Psychological States {\footnotesize (e.g., cognitive load, anxiety)} & 40.9\% & 10.8\% & 8.7\% & 39.0\% & 0.6\% \\
Interaction Quality {\footnotesize (e.g., user satisfaction, engagement)} & 38.7\% & 6.1\% & 8.2\% & 46.5\% & 0.5\% \\
\bottomrule
\end{tabularx}
\end{table*}

Relation type adds a further distinction. Mediation constitutes $8.2\%$ of claims in \emph{Technology Acceptance}, compared with $2.1\%$ in \emph{Performance}, while moderation accounts for $9.2\%$ in \emph{Technology Appraisals}. These distributions describe how relationships are represented within each area. The method profiles also differ. Controlled experiments account for $49.4\%$ of \emph{Performance} claims and $43.8\%$ of \emph{Technology Appraisals} claims. In \emph{Technology Acceptance}, $48.3\%$ are classified as uncontrolled observation and $17.7\%$ as controlled observation. These categories describe the recorded evidentiary basis, rather than rank methodological quality. The small theoretical category records claim-level coding within papers screened as empirical.

The direction of the reported effect is distinct from its human consequence. Within \emph{Psychological States}, $54.0\%$ of effect claims report increases, $30.8\%$ decreases, and $15.1\%$ null effects; the remaining $0.1\%$ have no recorded direction. In \emph{Performance}, the corresponding shares are $63.3\%$, $21.2\%$, and $15.4\%$, with $0.1\%$ unspecified. A decrease in a measure such as anxiety or completion time can be favorable, so these directional profiles must be read alongside the particular constructs and outcome coding. This distinction becomes especially relevant when several outcomes are reported for the same intervention.

\subsection{Connecting Findings Through Shared Constructs}

Following trust in both directions connects findings about system properties to findings about acceptance. The transparency--trust relationship comprises $121$ effect claims across $116$ paper records, while trust--intention to use comprises $57$ across $46$. Traversing the shared trust entity brings these bodies of evidence into the same inquiry. Each edge retains its own study contexts and sources, so the resulting path can be inspected as a connection among reported findings.

Mediation claims provide a further level of interpretation by recording pathways examined within a study. One retrieved example is Ngo's study of transparency in fake-news detection and social recommendations~\cite{Ngo2025}. The study reports that transparency relates to use intention through trust and certainty, alongside an inverted-U relationship between transparency and use intention. Read together, these findings qualify a simple account in which increasing transparency uniformly promotes adoption. Following the shared constructs leads from a common pairwise relationship to a more specific account of its operation. The nonlinear finding also illustrates why source inspection remains necessary when a directional edge compresses a more complex result.

\subsection{Examining Joint Outcomes Within Studies}

The study-arm links support examination of outcomes reported for the same intervention and study context. We joined effect claims by their canonical subject and source-linked study arm, then retrieved combinations of positively and negatively coded outcomes in different taxonomy categories. For \emph{AI assistance}, this query returned nine arms across nine paper records combining a positively coded \emph{Performance} finding with a negatively coded \emph{Psychological States} finding. These combinations identify cases for closer comparison; their interpretation depends on the outcomes and conditions represented by the individual claims.

An example retrieved around \emph{explanation} concerns the evaluation of TxGNN, a system for assessing drug-repurposing predictions~\cite{huang2024foundation}. In a study with twelve medical experts, explanations improved assessment accuracy and confidence while increasing the time spent evaluating predictions. The source describes the additional time as an opportunity to contextualize explanations using domain expertise. The joint profile therefore involves both evaluation effort and improved judgment. Keeping these findings connected avoids reducing the intervention to a single favorable or unfavorable outcome and helps readers examine which consequences matter for their intended use.

\subsection{Interpreting Relationships in Context}

An aggregate relationship can be examined at progressively narrower scopes. Among transparency--trust claims, $103$ are classified as uncontrolled observation, including $94$ increases, seven decreases, and two null findings. The fourteen claims classified as controlled experiments comprise eight increases and six null findings. These are distributions of extracted claims, and several claims can originate from the same paper. They describe how the retrieved evidence changes with a method filter, rather than estimate a pooled effect or establish an explanation for the difference.

Study context provides a more specific interpretation. In \emph{Guided By AI}, Ha et al.~\cite{ha2024guidedainavigatingtrust} examined four transparency conditions in visual data exploration. Transparency did not significantly change subjective trust in either the easy or hard task. Participants nevertheless used more suggestions in the harder task despite lower suggestion accuracy. Linking these findings brings task demands and observed reliance into an inquiry initially framed around transparency and trust. The relevant question becomes how people use the available guidance under particular conditions, extending beyond whether they report greater trust.

Together, these analyses move from the taxonomy and distributions of research across facets and areas to the joint outcomes, reported mechanisms, and study conditions behind individual relationships. The structured representation supports this progression by connecting findings across constructs while preserving access to the evidence needed to interpret each connection.

\section{User Study}

We conducted a task-based user study to understand how researchers use Atlas to explore literature, synthesize findings, and develop research questions. Our primary interest was qualitative, examining how participants navigated the evidence, which features supported their work, and where they encountered difficulties. Participants completed synthesis and open-ended exploration tasks using either Atlas or their customary research tools. Think-aloud recordings and post-task reflections provided insight into their approaches and experiences. We also collected expert ratings of the written syntheses to describe the overall pattern of output quality across conditions. These ratings provided additional context for interpreting participants’ approaches and experiences.

\subsection{Method}

We recruited 10 researchers and students in HCI and adjacent fields through university mailing lists. 5 participants used Atlas, and 5 used their customary research tools. The study was conducted under an IRB exemption protocol [redacted for anonymity].

Sessions were scheduled for 45 minutes and conducted online or in person according to participant preference. Sessions were recorded over Zoom, and participants were asked to think aloud throughout the tasks. Each session comprised three stages:

\begin{enumerate}
    \item \textbf{Pre-task survey}: Participants reported their academic or research background and the tools they ordinarily used to find and synthesize literature.
    \item \textbf{Synthesis and exploration tasks}: Participants synthesized evidence on two prescribed questions, then explored a topic of their choice to develop a research question. Participants in the Atlas condition first spent five minutes exploring the system.
    \item \textbf{Closing survey}: Participants reflected on their experience, including the support and difficulties they encountered during the tasks.
\end{enumerate}

The first task asked participants to write a one- to two-paragraph synthesis for each of two questions. A researcher uninvolved in the interface's development formulated the questions to reduce the risk of favouring Atlas:
 
\begin{enumerate}
    \item What is the impact of LLMs on critical thinking?
    \item What shapes users' trust in AI systems in healthcare, and what does that trust in turn affect?
\end{enumerate}

The second task asked participants to explore a topic of their choice and propose a research question motivated by the literature, such as an unexpected relationship or a disagreement worth investigating.

\subsection{Results}

\subsubsection{Task 1}

Two experts with SIGCHI reviewing experience independently rated the written syntheses, blind to condition. Adapting the rubric from~\cite{wang2024autosurvey}, we assessed three dimensions, including (1) Coverage, (2) Structure, and (3) Relevance, each scored from 1 to 5. Ratings were summarized descriptively to contextualize the qualitative analysis.

Across the three dimensions, syntheses produced with Atlas received a mean score of $3.03$ ($SD = 0.64$), compared with $2.87$ ($SD = 0.75$) for customary tools. Coverage showed the largest difference between Atlas ($M = 2.85$, $SD = 0.52$) and customary tools ($M = 2.50$, $SD = 0.79$). Relevance was also higher with Atlas ($M = 3.25$, $SD = 0.61$) than with customary tools ($M = 3.05$, $SD = 0.65$), while structure was similar between Atlas ($M = 3.00$, $SD = 0.90$) and customary tools ($M = 3.05$, $SD = 0.86$).

Given the small sample and brief exposure to Atlas, we interpret these differences descriptively and make no claim of statistical significance. Their direction is nevertheless consistent with Atlas’s aim of helping researchers locate and bring together relevant evidence.

\subsubsection{Task 2}

For this task, we examined how Atlas's support for top-down exploration and navigation between related concepts could help participants develop research questions, including through serendipitous encounters with unfamiliar findings. Participants explored a topic of their choice, looking for unexpected relationships, disagreements, or gaps that could motivate further investigation. Their responses illustrate several ways these encounters informed ideation. P4 followed a connection between Minecraft and AI literacy toward an interest in gamification for education; their reflection described discovering papers they otherwise would not have searched for. P5 questioned the boundaries of reported benefits, proposing a ``goldilocks'' window for AI backchanneling that considers when it becomes counterproductive and how this might differ across populations. P1 considered whether findings about cognitive engagement in coding and generative tasks extended to knowledge co-creation, then searched Google Scholar for additional studies. These accounts suggest that connections among findings can provide starting points for research ideas by introducing unfamiliar directions, prompting questions about the conditions of an effect, and guiding further searches.

\subsection{Qualitative Analysis}

This section examines the five Atlas participants' reflections on how the interface supported comparison, exploration, and synthesis. The authors developed, discussed, and agreed upon a codebook organized around the three design goals, with codes linked to the interface components presented in Section~\ref{sec:design}. We interpret the findings in relation to the literature informing these goals and identify opportunities for further iteration.

\subsubsection{Claim cards as a basis for comparison (DG1)}

Participants valued the common representation of findings presented through the claim cards. P2 described ``normalizing terms across papers'' as making studies ``very easy to compare,'' while P1 appreciated the ability to retrieve and group relevant citations. These accounts suggest that structured claims helped researchers recognize related findings across papers.

Participants also identified contextual information that could enrich this comparison. P1 requested clearer distinctions between paper types, quantitative and qualitative methods, and participant demographics. P2 valued the shared representation while remaining cautious about ``oversimplifying complex topics for the sake of comparison.'' Their comments suggest that recognizing a shared relationship is one part of comparison; examining differences between the contributing studies is another.

The claim card's evidence rail and source disclosure provide places to expose these distinctions. Consistent with recognition rather than recall~\cite{nielsen1994heuristics}, a further iteration could make study characteristics directly comparable across cards. This could also reduce the navigation overhead involved in reconstructing context across linked sources~\cite{conklin1987hypertext}.

P1 encountered apparently contradictory findings and questioned whether they reflected the literature or a system error. This suggests a further use for comparative evidence inspection. Bringing conflicting claims, source passages, and study contexts into a shared view could help researchers investigate the disagreement. The reflections establish an interest in this context, although they do not establish whether participants used source disclosures to verify claims.

\subsubsection{Concept pages as entry points for exploration (DG2)}

Participants valued the local Sankey diagrams on concept pages. P2 described the Sankey as helping them understand the problem space ``at a glance,'' while P4 particularly appreciated the smaller charts for showing related concepts. P5 similarly valued seeing connections together with their positive, negative, or neutral outcomes. These accounts are consistent with an overview supporting subsequent inspection of specific relationships~\cite{shneiderman1996eyes}.

Participants' experiences suggest that this progression also depends on reaching an appropriate starting scope. P5 described broadening a narrow search before navigating toward the intended topic, having to ``walk it back to then drill down via the UX.'' P2 found that sparse diagrams offered less value over customary tools. Progressive disclosure~\cite{nielsen2006progressive} could therefore extend to the transition between search and concept exploration. Connecting search results to related concepts and enclosing category pages could show readers how changing scope changes the available evidence.

The corpus graph offered a broader perspective but required additional guidance. P2 found it visually overwhelming and noted that it was not explained in the onboarding guide, while describing the rest of the interface as clear. Scaffolding directs attention toward consequential features of an unfamiliar task~\cite{wood1976tutoring}. A guided transition from a concept page into its position in the graph could introduce the larger structure through relationships the reader has already examined. Longer-term evaluation could establish how researchers' understanding of these views develops with experience.

\subsubsection{Collections and conversation within research workflows (DG3)}

The collection helped participants retain and organize discoveries. P4 reported that bookmarks focused their research scope ``without trying to hold a lot of things in my mind.'' They also encountered papers they ``otherwise would not have searched for.'' Their account connects the discovery of unfamiliar evidence with the practical work of retaining a relevant subset. Keeping that selection visible is consistent with reducing reliance on recall~\cite{nielsen1994heuristics}.

The conversational pane supported another aspect of exploration. P3 valued seeing ``the intersection of different ideas'' and described the assistant as their favourite part of the system. These experiences reflect the learning and investigation involved in exploratory search~\cite{marchionini2006exploratory}. Finding relevant evidence provides material for further interpretation as the researcher's question develops. Closer coordination between Sankey selections, saved claims, and scoped conversations could support this progression. P3's preference also motivates examining how the visual and conversational components contribute individually and together.

Participants situated Atlas within existing research workflows. P1 anticipated using it for initial orientation and examining general trends, while continuing to use digital libraries for specialized searches. P4 and P5 similarly identified early exploration and ideation as promising uses. These intended uses position Atlas within an evolving research process~\cite{marchionini2006exploratory}. Exporting selected claims and citations could preserve the results of exploration as researchers move into focused searching, reading, and writing.

P1 also wondered whether abundant evidence around an initial concept, such as trust, might discourage broader exploration. This suggests an opportunity for scaffolding that offers further directions while preserving the reader's control~\cite{wood1976tutoring,nielsen1994heuristics}. The category pages and corpus views could provide invitations to inspect an enclosing category or neighbouring concepts while retaining the current collection and line of inquiry.

\section{Discussion}
\subsection{Implications for Research and Practice}

Section~\ref{sec:intro-gap} describes a growing difficulty in establishing what is
already known across fragmented literatures. Finding-level synthesis offers
a way to make connections across studies whose shared concerns may be
obscured by differences in terminology or disciplinary framing. Participants'
appreciation of normalized claims and concept-level overviews suggests how
this representation can support initial orientation and comparison. For
researchers, these connections could help bring unfamiliar work into view,
examine apparent disagreements, and situate new questions within accumulated
evidence. Preserving study context is central to this process, since similar
reported relationships may arise under different methods, populations, and
conditions.

The same difficulty extends to people drawing on research to inform design
and policy. Designers and practitioners could use connected findings to
examine how outcomes vary across systems and settings relevant to their work.
Policymakers and evidence reviewers could explore findings around a question,
inspect which populations and contexts are represented, and follow aggregate
patterns back to their sources. These potential uses respond to the pressures
on evidence synthesis outlined in the introduction by making a body of
research easier to inspect and compare. Although our study focused on
researchers, it motivates investigating how structured synthesis could
support these broader audiences and the judgments their work requires.

\subsection{Generalizability of the Synthesis Approach and Visualization}

The pipeline provides a foundation for extending finding-level synthesis to
other empirical literatures facing similar challenges of scale and
fragmentation. Its stages can be adapted through domain-specific screening
criteria, claim schemas, and facets, while the interface offers a reusable
pattern for moving between broad categories, individual findings, and source
evidence. Collaboration with domain experts could guide this adaptation by
identifying relevant forms of evidence, refining shared terminology, and
shaping meaningful categories. Iterative application across diverse paper
samples could progressively enrich the structured extraction and expand the
range of findings it captures.

The current Atlas's emphasis on relational findings also opens opportunities
for complementary forms of synthesis. Comparative and attribute claims could
support views that bring together differences across conditions and properties
of constructs, enriching the context surrounding the relation graph. Future
work could extend extraction and visualization to descriptive and conceptual
literature, helping readers compare accounts, explore frameworks, and connect
conceptual propositions to empirical findings. These extensions could bring
a wider range of research contributions into a shared space for exploration
while preserving their distinct forms and connections to source material.

\subsection{The Future of Research Sensemaking and Knowledge Synthesis}

The growth described in the introduction calls for ways of understanding
research that can develop alongside the literature itself. The structured
synthesis underlying Atlas could inform digital libraries in which readers
navigate accumulated findings alongside documents. Our study suggests
complementary roles for visual exploration, evidence collection, and
conversation as researchers develop their questions. Connecting these
activities could help future generations of researchers find their footing
in unfamiliar literatures and carry discoveries into sustained reading and
writing through annotation, export, and integration with reference managers.

Incremental ingestion provides a basis for this infrastructure to evolve.
As findings accumulate around existing constructs, tracking changes in
claims, terminology, and taxonomies could help readers examine how an
evidence base develops and revisit earlier syntheses in light of new work.
This direction returns to Engelbart's vision of augmenting human intellect.
AI-assisted organization could make a growing literature more available for
human interpretation, helping researchers inspect connections, question
emerging patterns, and build on prior knowledge.

\subsection{Ethical Considerations}
Our corpus is limited to open-access scholarly literature and is used solely for academic research. We contacted the relevant publishers to request permission for full-text acquisition and processing. 

\subsection{Limitations and Future Work}
\label{sec:limitations}

The current corpus provides a starting point for progressively broader
coverage. Its scope is shaped by the metadata query, screening criteria,
and full-text availability. Future iterations could expand retrieval sources
and assess screening recall against expert-curated samples, bringing more
venues and forms of empirical research into the Atlas. Making coverage
visible across populations, contexts, and publication sources could also
help readers distinguish areas with little represented evidence from
opportunities requiring further investigation.

Further validation could strengthen the representation and its interpretation.
Expert assessment of extracted claims, entity mappings, and taxonomy
assignments could identify where additional context or finer distinctions
would improve comparison. The existing links to source material provide a
basis for incorporating corrections into this process. Richer summaries of
study design and evidentiary support could also help readers interpret
aggregate patterns, since claim frequency alone does not establish evidence
strength, repeated relationships do not necessarily indicate replication,
and observational findings do not establish causality.

The user study offers initial insights that could guide evaluation over
longer periods and across broader audiences. With ten participants and brief
exposure to Atlas, the findings describe early experiences rather than
sustained research outcomes. Longitudinal studies could examine how readers
verify claims, investigate disagreements, and carry discoveries into their
research, while assessing how visual and conversational features contribute
to these activities. Extending evaluation to practitioners and policymakers
could inform adaptations to their questions, evidence needs, and workflows.

\bibliographystyle{ACM-Reference-Format}
\bibliography{references}

\appendix

\section{OpenAlex OQL Query}
\begin{lstlisting}[literate={→}{$\rightarrow$}1 {↑}{$\newcheckmark$}1 {↓}{$\newcrossmark$}1]]
works where title/abstract has (
    ("human" or "user" or "participant" or "team" or "people")
    and (
      "ai" or "llm" or "artificial intelligence" or "machine learning" or "chatbot" or "conversational agent" or "large language model" or "recommender system" or "recommendation system"
    )
  )
  and DOI prefix is (
        10.1001 or 10.1002 or 10.1007 or 10.1016 or 10.1017 or 10.1038 or 10.1051 or 10.1055 or 10.1057 or 10.1075 or 10.1080 or 10.1088 or 10.1093 or 10.1098 or 10.1101 or 10.1108 or 10.1109 or 10.1111 or 10.1145 or 10.1167 or 10.1177 or 10.1186 or 10.1192 or 10.1287 or 10.1371 or 10.1515 or 10.1609 or 10.1613 or 10.17605 or 10.17613 or 10.18653 or 10.20944 or 10.2196 or 10.22318 or 10.24251 or 10.2478 or 10.24963 or 10.25300 or 10.2991 or 10.31219 or 10.31234 or 10.3389 or 10.3390 or 10.36227 or 10.3765 or 10.4018 or 10.4324 or 10.48550
  )
  and (open access is (true))
  \end{lstlisting}

  \section{User Study Task Prompts}
\label{appendix:study-tasks}

\subsection{Getting Familiar (Atlas Condition Only)}

Open the Atlas of Human-AI Interaction [Link redacted for anonymity]. The Atlas is a research sensemaking tool that synthesizes over 32,000 papers in human-AI interaction into an interactive knowledge graph. Each paper is extracted into structured claims, linking concepts (e.g. "transparency increases trust") together, so you can see the larger pattern they form. Rather than searching for papers, you search for a concept and see what the literature says it affects, and what affects it.

What to do
\begin{itemize}
    \item Look at the Learn More page (the link under the search bar in the landing page)
    \item Click whatever looks interesting. Follow whatever you are curious about. If something is confusing, say so out loud and keep going.
\end{itemize}

Explore the 'Finding' listing view
\begin{itemize}
    \item Click on "trust" to see research findings related to trust
    \item Expand some claims to view detailed statistics and specific clauses
    \item Use the filter on the left to sort findings by demographic and system type
    \item Click the diagram at the top to filter findings by related concepts
    \item Check related topics at the bottom of the page
\end{itemize}

Try the LLM feature
\begin{itemize}
\item Bookmark claims by clicking the Bookmark icon at the top right of each claim card
\item Visit the Bookmark tab and prompt the LLM to analyze your saved claims.
\end{itemize}

A other few things you might try, in any order:
\begin{itemize}
\item Type something you know about into the search bar 
\item Click one of the example topics
\item Open a concept and look at what it connects to
\end{itemize}

Please keep talking while you do this. Say what you expected, what you found, and what you think each screen is showing you.

\subsection{Task 1: Literature Synthesis}

Researchers often need to work out what a body of literature actually says about a topic, whether that's scoping a new project or writing a related-work section. For each question below, please write key insights that could serve as a background and related work section in a research paper, for example, what the findings in the field are, how they relate to each other, and where the research are currently exploring. We don't expect full comprehensiveness in seven minutes, cover what you can in the time. The Next button will appear after 12 minutes. 

\begin{itemize}
    \item 1-2 paragraphs for each question as you would for a knowledgeable colleague.
    \item Be thorough: Give an account of the topic as a whole, not just the single strongest finding you come across.
    \item Be precise: Name concepts as the literature names them, cite what you draw on, and make clear how much evidence stands behind each point you make.
\end{itemize}

\begin{enumerate}
    \item What is the impact of LLMs on critical thinking?
    \item What shapes users' trust in AI systems in healthcare,
    and what does that trust in turn affect?
\end{enumerate}

\subsection{Task 2: Creating Research Questions}

Reading across a body of literature often turns up a question worth asking rather than an answer. This task asks you to look for one. The Next button will appear after 8 minutes.

Explore whatever interests you: your own area, something you noticed in Task 2, or anything else worth a closer look.

You're looking for something like an unexpected relationship, a disagreement worth resolving, a gap where you'd expect more work, or a pattern that suggests an untested mechanism.

\end{document}